\documentclass[]{aa} 

\usepackage{graphicx}
\usepackage{txfonts}
\usepackage{natbib}
\usepackage{longtable}

\begin{document}

\title{Estimation of  high-resolution dust column density maps}
\subtitle{Comparison of modified black-body fits and radiative
transfer modelling}

\author{M.     Juvela\inst{1},
J.     Malinen\inst{1},
T.     Lunttila\inst{1}
}

\institute{
Department of Physics, P.O.Box 64, FI-00014, University of Helsinki,
Finland, {\em mika.juvela@helsinki.fi}
}

\authorrunning{M. Juvela et al.}

\date{Received September 15, 1996; accepted March 16, 1997}

\abstract
% context heading (optional)
% {} leave it empty if necessary  
{
Sub-millimetre dust emission is routinely used to derive the column
density $N$ of dense interstellar clouds.  The observations consist of
data at several wavelengths but also, with increasing wavelength, of
poorer resolution.  Procedures have been proposed for deriving higher
resolution maps of $N$. In this paper the main ones are called Methods
A and B.
Method A uses low-resolution temperature estimates combined with
higher resolution intensity data. Method B is a combination of column
density estimates obtained using different wavelength ranges.
}
% aims heading (mandatory)
{
Our aim is to determine the accuracy of the proposed methods relative
to the true column densities and to the estimates that could be
obtained with radiative transfer modelling.
}
% methods heading (mandatory)
{
We used magnetohydrodynamical (MHD) simulations and radiative transfer
calculations to simulate sub-millimetre surface brightness
observations at the wavelengths of the {\em Herschel Space
Observatory}. The synthetic observations were analysed with the
proposed methods and the results compared to the true column densities
and to the results obtained with simple 3D radiative transfer
modelling of the observations.
}
% results heading (mandatory)
{
Both methods give relatively reliable column density estimates at the
resolution of 250\,$\mu$m data while also making use of the longer
wavelengths. In case of high signal-to-noise data, the results of
Method B are better correlated with the true column density, while
Method A is less sensitive to noise. When the cloud has internal
heating sources, Method B gives results that are consistent with those
that would be obtained if high-resolution data were available at all
wavelengths. Because of line-of-sight temperature variations, these
underestimate the true column density, and because of a favourable
cancellation of errors, Method A can sometimes give more correct
values. Radiative transfer modelling, even with very simple 3D cloud
models, usually provides more accurate results. However, the
complexity of the models that are required for improved results
increases rapidly with the complexity and opacity of the clouds.
}
% conclusions heading (optional), leave it empty if necessary 
{
Method B provides reliable estimates of the column density, although
in the case of internal heating, Method A can be less biased
because of fortuitous cancellation of errors. For clouds with a
simple density structure, improved column density estimates can be
obtained even with simple radiative transfer modelling.
}
\keywords{
ISM: clouds -- Infrared: ISM -- Radiative transfer -- Submillimeter: ISM
}

\maketitle
%
%________________________________________________________________

\section{Introduction}

Sub-millimetre and millimetre dust emission data are widely used to
map the structure of dense interstellar clouds \citep{Motte1998,
Andre2000}. The large surveys of {\em Herschel} are currently
providing data on, for example, major nearby molecular clouds
\citep{Andre2010} and on the whole Galactic plane
\citet{Molinari2010}. The column densities derived from the emission
depend not only on the strength of the emission but also on its
spectrum. By covering the far-infrared part of the spectrum, {\em
Herschel} observations are also sensitive to the dust temperature.
Accurate estimates of the column density require accurate values of
both temperature and dust opacity. In addition to being a tracer of
cloud structure, the dust emission also carries information on the
properties of the dust grains themselves. The properties are observed
to change between clouds, which is associated with differences in
the grain optical properties, the size distributions, the presence of
ice mantles, and possibly even temperature-dependent optical
characteristics \citep{Ossenkopf1994,Stepnik2003, Meny2007,
Compiegne2011}. 

Measurements of dust emission can be complemented with information
from other tracers. Both dust extinction and scattering can be
observed using near-infrared (NIR) wavelengths. These would be
valuable because they are independent of that dust temperature that is a
major uncertainty in the interpretation of emission data.  Additional
data would also help constrain the dust properties; see, e.g.,
\citet{Goodman2009} and \citet{Malinen2012} for a comparison of the
use of dust emission and extinction. Dust extinction has been mapped
over large areas \citep{Lombardi2006, Goodman2009, Schneider2011}, but
high resolution observations are expensive and restricted to small
fields. The same applies to observations of NIR scattered light that
still exist only for a few clouds \citep{Lehtinen1996, Nakajima2003,
Foster2006, Juvela2008, Nakajima2008, Malinen2013}. Measurements of
scattered light are even rarer in the mid-infrared, and they
probably tell more about the dust grains than the column density
\citep{Steinacker2010, Pagani2010}. Therefore, in most cases one must
rely on correct interpretation of the dust emission, preferably at
far-infrared  and longer wavelengths. Below 100\,$\mu$m the situation
is complicated by the contribution of transiently heated grains and
by the sensitivity to the shorter wavelengths, for which the clouds
are typically optically thick.

In addition to the uncertainty of the intrinsic grain properties, 
interpretation of dust emission data is affected by two main problems,
the effect of noise and the effect temperature variations. The noise
is particularly problematic if one tries to determine both the dust
temperature and the dust emissivity spectral index $\beta$
\citep{Shetty2009b, Juvela2012_chi2}. Therefore, most estimates of
cloud masses are derived by assuming a constant value of $\beta$.
However, there is always dust with different temperatures along the
line-of-sight and at different positions within the beam, and
this affects the mass estimates \citep{Shetty2009a, Malinen2011,
Juvela2012_TB, YsardJuvela2012}. Because the emission of warm dust is
stronger than the emission of colder dust of the same mass, the colour
temperature derived from the observed intensities overestimates the
mass-averaged temperature. The greater the temperature variations are,
the more the dust mass is underestimated \citep{Evans2001,
StamatellosWhitworth2003, Malinen2011, YsardJuvela2012}. The problem
could be solved only if the temperature structure of the source were
known so that the effect could be determined with modelling.

One would like to measure the density and temperature structure of the
clouds not only as reliably as possible but also with as high a
resolution as possible. The resolution depends on the telescope and
the wavelengths used. For {\em Herschel}, the resolution varies from
less than 8$\arcsec$ at 100\,$\mu$m to $\sim$37$\arcsec$ at
500\,$\mu$m. The standard way to calculate a column density map is to
convert all data first to the lowest common resolution. Thus, most of
the input data has significantly higher resolution than the result.
Therefore, it would be beneficial to find ways to combine the
data in a way that, although all wavelengths are used, the final map
would retain a resolution better than that of the longest waveband.

\citet{Juvela2012_GCC_III} examined cloud filaments using column
densities derived from {\em Herschel} 250\,$\mu$m surface brightness
data at $\sim$20$\arcsec$ resolution and dust colour temperature at
$\sim$40$\arcsec$ resolution. It was argued that the effective
resolution must be better than 40\,$\arcsec$ because, on small
scales, the temperature changes are small.  \citet{Palmeirim2013}
presented a better justified method that combined {\em Herschel} data at
160, 250, 350, and 500\,$\mu$m to produce high resolution column
density maps. The methods may raise the question, what is the actual
resolution of the maps. Furthermore, if one uses temperature maps of
different resolution, they will be affected differently by the
line-of-sight temperature variations and this could be reflected in
the results. The aim of this paper is to investigate these questions.
In the work we use the results of radiative transfer calculations,
where the model clouds are the result of magnetohydrodynamical (MHD)
simulations and the clouds may also contain point sources that
introduce strong local temperature gradients. We compare the results
of the above mentioned and similar methods (see
Sect.~\ref{sect:methods}) with the true column densities known from the
models. We also examine the accuracy to which the column densities
can be determined by carrying out radiative transfer modelling of the
data. Such modelling has been applied, also recently, in the
examination of the density and temperature structure of dense clouds
\citep[e.g.][]{Ridderstad2010, Juvela2012_CrA_III, Nielbock2012,
Wilcock2012}.

The content of the paper is the following. In Sect.~\ref{sect:simu} we
present the cloud models and the calculation of the surface brightness
maps. In Sect.~\ref{sect:methods} we describe the basic estimation of
the colour temperatures and column densities and present the two
methods that are used to convert surface brightness data back to
high-resolution column density maps. In Sect.~\ref{sect:results} we
present the main results, comparing the column density estimates to
the true column densities in the cloud models and to the column
densities that would be recovered by higher resolution observations.
In Sect.~\ref{sect:modelling} we construct three-dimensional models
that are adjusted to reproduce the surface brightness observations and
in this way used to estimate the column densities. The column densities
of these models are again compared to the column density in the
original model cloud. The results are discussed in
Sect.~\ref{sect:discussion} where we also draw the final conclusions
on the relative merits of the methods used.

\section{Simulated observations} \label{sect:simu}

We use surface brightness maps calculated for two MHD models that are described
in more detail in \citet{Malinen2011} and \citet{Juvela2012_mhdfil}.
Cloud I corresponds to an isothermal magnetohydrodynamical (MHD) simulation
carried out on a regular grid of 1000$^3$ cells \citep{PadoanNordlund2011}. The
calculations included self-gravity and the snapshot corresponds to situation
before any significant core collapse. The cloud is scaled to a linear size of
6\,pc and a mean density of $n$(H)=222.0\,cm$^{-3}$, giving a mean visual
extinction of 2$^{\rm m}$. The model is the same that was used in
\cite{Juvela2012_mhdfil} to study filamentary structures.
Cloud II was calculated using the adaptive mesh refinement (AMR) code Enzo
\citep{Collins2010}. The model has a base grid of 128$^3$, four levels of
refinement, and an effective resolution of 2048$^3$ cells. The model has been
discussed in \cite{Collins2011} and in \cite{Malinen2011} (called Model II in
that paper). As in the case of Cloud I, the MHD calculations assumed an
isothermal equation of state. The linear size and the mean density of the model
are scaled to 10 pc and 400 cm$^{-3}$. This gives an average column density of
$N$(H)=1.23$\times$10$^{22}$\,cm$^{-2}$ that corresponds to
$A_{V}\sim$6.6$^{\rm mag}$. The extinction reaches 20$^{\rm m}$ in less than
2\% of the map pixels. 

For the radiative transfer modelling, the density fields were resampled onto
hierarchical grids. The gridding preserves the full resolution in the dense
parts of the model clouds but, to speed up the calculations, the resolution is
degraded in the low density regions. Occasionally the greater size of some cells
along the line-of-sight produces noticeable artefacts in the surface brightness
maps which, however, usually disappear when the data are convolved with the
telescope beam. Because the parameters that are being compared (i.e., the true
and the estimated column densities) refer to the same discretisation, possible
discretisation errors do not directly affect this comparison.

The dust temperature distributions and the emerging dust continuum
emission were calculated with the radiative transfer code described in
\cite{Lunttila2012}. The clouds are illuminated externally by an
isotropic interstellar radiation field \citep{Mathis1983} and the dust
properties correspond to those of the normal diffuse interstellar
medium \citep{Draine2003} with a gas-to-dust ratio of 124 and
$R_{V}$=3.1. The calculations are described in more detail in
\citet{Malinen2011} and \citet{Juvela2012_mhdfil}. We refer to the
densest sub-structures of the model clouds as cores. We will also
examine a case where the cores of Cloud II, which already are known to
be gravitationally bound, have internal heating sources. The
properties of the sources and the procedures used in their modelling
are described in \cite{Malinen2011}.  There are 34 sources with
luminosities between 2.1 and 82 solar luminosities. For the present
study, their main effect is how they modify the three-dimensional
distribution of dust temperature and how that is reflected in the
surface brightness measurements. With the assumed cloud distance of 500\,pc,
the sources can locally raise the dust colour temperature to 20--30\,K
when observed at the resolution of 40$\arcsec$.

We use the radiative transfer modelling to simulate observations by the {\em
Herschel Space Observatory} \citep{Pilbratt2010}. The calculations result in
synthetic surface brightness maps at 160, 250, 350, and 500\,$\mu$m. The map
size is 1000$\times$1000 pixels for Cloud I and 2048$\times$2048 pixels for
Cloud II. As a default we assume noise levels of 3.7, 1.20, 0.85, and
0.35\,MJy\,sr$^{-1}$ per beam for 160, 250, 350, and 500\,$\mu$m, respectively.
However, we also examine cases with noise 0.3 or 3.0 times these values.The
pixel size of the maps is set to a value of 2.0$\arcsec$. During the analysis
the maps are convolved with the assumed beam sizes of 12.0$\arcsec$,
18.3$\arcsec$, 24.9$\arcsec$, and 36.3$\arcsec$, for the four bands in the
order of increasing wavelength. The values correspond to the approximate beam
sizes of {\em Herschel} \citep{Poglitsch2010, Griffin2010}.

\begin{table*}
\caption{A summary of the methods used for the estimation of column density.}
\label{table:methods}
\begin{tabular}{ll}
\hline
Name    &   Description   \\
\hline
default &  Modified black body fitting with single resolution data, Eq.~\ref{eq:colden} \\
A       &  $T_{\rm dust}$ at 36.3$\arcsec$ and $I_{\nu}(250\mu{\rm m})$ at 18.3$\arcsec$;
           Sect.~\ref{sect:methodA}, \cite{Juvela2012_GCC_III} \\
B       &  $N(500) + \left[ N(350) - N(350\rightarrow 500) \right]
           +  \left[ N(250) - N(250\rightarrow 350) \right ]$,
           Eq.~\ref{eq:palmeirim}, \cite{Palmeirim2013}$^{1,2}$  \\
C       &  $p_1 \times N(500) + p_2 \times \left[ N(350) - N(350\rightarrow 500) \right]
           + p_3 \times \left[ N(250) - N(250\rightarrow 350) \right]$; see
           Sect.~\ref{sect:methodC} \\
D       &  $p_1 \times N(250 \rightarrow 500) + p_2 \times N(350 \rightarrow 500) + p_3 \times N(500)$ ; see
           Sect.~\ref{sect:methodC} \\
NL      &  $p_1 \times N(250 \rightarrow 500) + p_2 \times N(350 \rightarrow 500) 
           + p_3 \times N(500) +
           p_4 \times \left[ N(500)-N(250 \rightarrow 500) \right]^2$; see Sect.~\ref{sect:methodC} \\
RT      &  radiative transfer modelling of surface brightness data; Sect.~\ref{sect:RT} \\
\hline
\end{tabular}
\\
$^1$\,$N(\lambda)$ is the column density estimate obtained with surface brightness data 
at wavelengths from 160\,$\mu$m to $\lambda$, at the resolution of the longest wavelength. \\
$^2$\,$N(\lambda_1 \rightarrow \lambda_2)$ is $N(\lambda_1)$ convolved to the resolution of 
observations at wavelength $\lambda_2$.\\
\end{table*}

\section{Analysis methods} \label{sect:methods}

In this section, we present the methods that are used to estimate the
dust temperature and column density without resorting to full
radiative transfer modelling. In particular, we recount the procedures
used in \citet{Juvela2012_GCC_III} and \citet{Palmeirim2013} to
increase the spatial resolution of the column density maps.
The methods are explained below and a summary of all the analytical
combinations of individual column density estimates are summarised in
Table~\ref{table:methods}.

\subsection{Estimation of column density} \label{sect:colden}

The basic principles of the column density estimation are the same for
all methods. The observed intensity $I_{\nu}$ is approximated with a
modified black body curve
\begin{equation}
I_{\nu} = B_{\nu}(T) (1-e^{-\tau}) \approx B_{\nu}(T) \tau =
B_{\nu}(T) \kappa_{\nu} N.
\label{eq:colden}
\end{equation}
The equation assumes that the medium can be described with a single
temperature value. The included approximation of the exponential term
is valid if the optical depth $\tau$ is much smaller than one. This
is the case for the models and the wavelengths examined in this
paper. The optical depth $\tau$ is the product of dust opacity at the
frequency in question, $\kappa_{\nu}$, and the column density $N$.
Thus, the equation can be used to estimate the column density,
provided that the dust temperature $T$ is known. If we assume for the
opacity a frequency dependence of $\kappa_{\nu} \propto \nu^{\beta}$
with some fixed value of the emissivity spectral index $\beta$, the
value of $T$ can be estimated with observations of two or more
wavelengths, the latter requiring a fit to the observed intensities.
We carry out these as least squares fits. Whenever the source
contains temperature variations, the colour temperature obtained from
these fits is only an approximation of the mass averaged dust
temperature \citep{Shetty2009a, Malinen2011, Juvela2012_TB,
YsardJuvela2012}. This is one of the main reason why the morphology
of the derived column density maps (i.e., column density contrasts)
deviate from the reality.

The least squares fits are carried out pixel by pixel, weighting the
data points according to the observational noise. The fitted 
temperature and intensity are inserted to Eq.~\ref{eq:colden} for the
calculation of an estimate of $N$. If data are available at more than
two wavelengths, also the dust emissivity spectral index $\beta$ could
be determined. However, in this paper the value of $\beta$ is kept
fixed to the value of 2.0. In the dust model used in the radiative
transfer calculations, the spectral index changes only a little as a
function of wavelength and is $\sim$2.08 between the wavelengths of
160\,$\mu$m and 500\,$\mu$m. 

In the case of real observations, the absolute value of the opacity
$\kappa_{\nu}$ is a major source of uncertainty. In this paper, we are
not interested in this factor and simply scale the median of the 
estimated column density maps to the median of the true column density
that is known for the model clouds.

\subsection{Higher resolution estimates: Method A} \label{sect:methodA}

\cite{Juvela2012_GCC_III} used {\em Herschel} observations to estimate the
column density in the usual way, convolving all surface brightness
data to the resolution of the 500\,$\mu$m data. However, the paper
also used alternative column density estimates that were obtained
combining the colour temperatures at the 40$\arcsec$ resolution with
250\,$\mu$m surface brightness data at a resolution of 20$\arcsec$. 
It was argued that the effective resolution of those maps would be
close to 20$\arcsec$ because the variations in the surface brightness
are stronger than the effects of colour temperature variations. It is
not clear to what extent this is correct. This also depends on the
difference between the colour temperature and the true mass averaged
dust temperature. For example, a compact cold core can have a
significantly lower physical temperature without a significant effect
on the colour temperature that is dominated by emission from warmer
regions. This means that although lower resolution of the temperature
map does increase errors, these may not always be very significant.

\subsection{Higher resolution estimates: Method B} \label{sect:methodB}

In \cite{Palmeirim2013} a higher resolution column density map was
obtained combining column density maps that were derived using
different sets of wavelengths. The data consisted of {\em Herschel} at
160, 250, 350, and 500\,$\mu$m. One starts by calculating column
density maps $N(250)$, $N(350)$, and $N(500)$ that are based on data
up to the specified wavelength and convolved to the corresponding
resolution. For example, $N(350)$ is based on the 160\,$\mu$m,
250\,$\mu$m, and 350\,$\mu$m maps that are convolved to the resolution
of the 350\,$\mu$m map, $\sim$25$\arcsec$. If one convolves such a
column density map to lower resolution, one also obtains estimates for
the difference in the structures that are visible in the two versions.
We use the notation $N(\lambda_1 \rightarrow \lambda_2)$ to denote a
column density map that is first estimated using data at wavelengths
$\lambda \le \lambda_1$ and at the resolution of the observations at
$\lambda_1$ and is then convolved to the resolution of observations at
wavelength $\lambda_2$.
The final estimate of the column densities is obtained as a
combination
\begin{eqnarray}
N & = & N(500) +  \left[ N(350) - N(350\rightarrow 500) \right] 
 \nonumber
\\
  &   & +  \left[ N(250) - N(250\rightarrow 350) \right].
\label{eq:palmeirim}           
\end{eqnarray}
$N(500)$ is the best estimate of column density at low resolution. The
other terms add information on structures that are visible at the
resolution of 350\,$\mu$m data but not at the resolution of
500\,$\mu$m data and finally the structures that are visible at
250\,$\mu$m but not at the resolution of the 350\,$\mu$m data. In
principle, the method thus provides estimates for the column density at
the resolution of the 250\,$\mu$m data, $\sim 18 \arcsec$.

The estimates $N(250)$, $N(350)$, and $N(500)$ will be different and
not only because of the different resolution. By using different sets
of wavelengths, one will not only have different noise levels but also
the bias of each estimates will be different \citep{Shetty2009a,
Shetty2009b, Malinen2011}. The biases are related to the temperature
distribution of the source. In particular, without data at long
wavelengths, one will be relatively insensitive to very cold dust. Of
course, if the estimates were identical, one could use directly the
$N(250)$ map. With Eq.~\ref{eq:palmeirim}, one can include all the
data although, of course, the correction terms $\left[ N(350) -
N(350\rightarrow 500) \right]$ and $\left[ N(250) - N(250\rightarrow
350) \right]$ (i.e., estimates of small-scale structures) will be
progressively more insensitive to cold emission.

\subsection{Other higher resolution estimates} \label{sect:methodC}

Because the bias of the column density estimates depends on the
wavelengths used, it is possible that this particular combination of
$N(250)$, $N(350)$, and $N(500)$ is not optimal for the overall
accuracy of the results. Therefore, we will later in
Sect.~\ref{sect:results} also examine some other linear and non-linear
combinations. In particular, we will examine which linear combination
of the terms $N(500)$, $\left[ N(350) - N(350\rightarrow 500)
\right]$, $\left[ N(250) - N(250\rightarrow 350) \right]$ gives the
best correlation with the true column density. This will be called the
Method C although, of course, in the case of real observations the
values of these coefficients could not be estimated. Finally, to
examine the effects of the wavelength-dependent biases further, we
consider a linear combination of $N(250 \rightarrow 500)$, $N(350
\rightarrow 500)$, and $N(500)$ where all surface brightness maps are
first converted to the resolution of $N(500)$. We call this Method D.
Because already all the surface brightness data are convolved to a
common resolution, Method D does not try to improve the spatial
resolution of the column density map, only the correctness of the low
resolution estimates. 

The final analytical method includes one additional, non-linear term.
The method is denoted by NL and, in addition to the terms
already included in Method D, it contains a term $\left[ N(500)-N(250)
\right]^2$. At this point this is examined only as one possible idea
of taking the non-linear features introduced by the 
colour temperature bias into account.

\subsection{Radiative transfer modelling} \label{sect:RT}

As an alternative to the ``analytical'' Methods A and B, we will also
examine radiative transfer modelling as a tool for the column density
determination. We construct a simple three-dimensional model for a
source and carry out radiative transfer modelling to
predict the dust emission at the observed wavelengths. The model is
then adjusted until there is a satisfactory correspondence between
the observed and the modelled intensities. The column density
estimates are then read from the final model cloud. 

In practice, we construct three-dimensional models where the density
distribution is discretised onto a grid of 81$^3$ cells. The radiative
transfer calculations are carried out with a Monte Carlo Program
\citep{Juvela2003, Juvela2005} using the same dust model as in the
original calculations that were used to create the synthetic
observations of Cloud I and Cloud II. Therefore, we exclude from
consideration the errors that would be caused by wrong assumptions of
dust opacity (and, consequently, of the dust emissivity spectral
index). These errors are not included in the results of Methods A and B
either. However, one must note that in the modelling we assume
consistency not only at the sub-millimetre wavelengths but also at the
short wavelengths where the dust grains are absorbing most of their
energy. As the external radiation field we also used the same
\cite{Mathis1983} model as in the original simulations but absolute
level of the radiation field is not assumed to be known. 

The cell size of the constructed 81$^3$ cell models is set equal to
4$\arcsec$. When the observations and the model are compared, the
160\,$\mu$m and 250\,$\mu$m data are convolved to the resolution of
18.3$\arcsec$ and the 350\,$\mu$m and 500\,$\mu$m data to the
resolution of the observations, 24.9$\arcsec$ and 36.3$\arcsec$,
respectively. The details of the assumed density distribution and the
procedures used to update the model clouds are discussed in more
detail in Sect.~\ref{sect:modelling}. However, because the column
densities are adjusted directly based on the 250\,$\mu$m data, the
modelling should also provide column density estimates at the same
resolution.

\section{Results} \label{sect:results}

\subsection{Comparison of Methods A and B} \label{sect:comparison}

Using the surface brightness maps at 160, 250, 350, and 500\,$\mu$m
resulting from the radiative transfer modelling
(Sect.~\ref{sect:simu}) we calculate column density maps using the
Methods A and B (see Sect.~\ref{sect:methods}). The results are
compared with the true column density of the model clouds and with
column density maps that are derived from synthetic observations with
a uniform spatial resolution of 18.3\,$\arcsec$ and without 
observational noise.

\begin{figure}
\centering
\includegraphics[width=8.7cm]{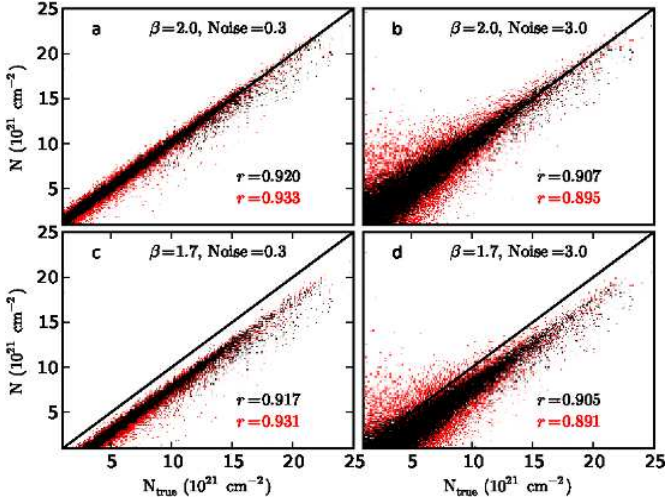}
\caption{
Column densities estimated with Method A (black points) and Method B
(red points) versus the true column density of Cloud I. The frames
correspond to 0.3 or 3.0 times the default noise (see text).  The
numbers indicate the correlation coefficients $r$ for Method A and Method
B, respectively, for data with $N_{\rm true}> 5\times
10^{21}$\,cm$^{-2}$. For illustration, we have included in the lower
frames the corresponding results obtained with a spectral index value
of $\beta=1.7$ instead of $\beta=2.0$.
}
\label{fig:tff031_colden}%
\end{figure}

\begin{figure}
\centering
\includegraphics[width=8.7cm]{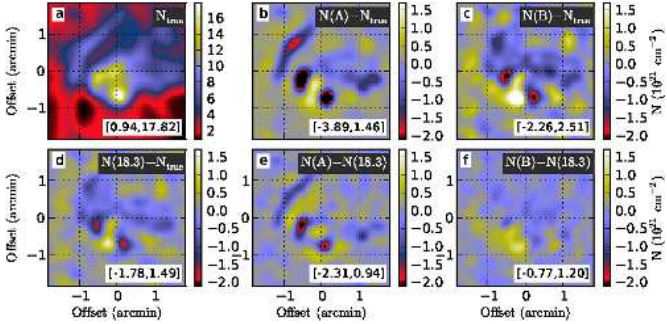}
\caption{
Column density maps for a selected small region in Cloud I. Frame $a$
shows the true column density of the model cloud and the frames $b$
and $c$ the errors in the estimates of Method A and Method B,
respectively ($N({\rm A})-N_{\rm true}$ and $N({\rm B})-N_{\rm
true}$). Frame $d$ shows the column density errors when estimates are
calculated using data with 18.3$\arcsec$ resolution at all
wavelengths, $N(18.3)-N_{\rm true}$. The frames $e$--$f$ show the
errors of Method A and Method B relative to the estimates from
18.3$\arcsec$ resolution data. The range of data values is given at the
bottom of each frame. 
}
\label{fig:tff031_map}%
\end{figure}

\begin{figure}
\centering
\includegraphics[width=8.7cm]{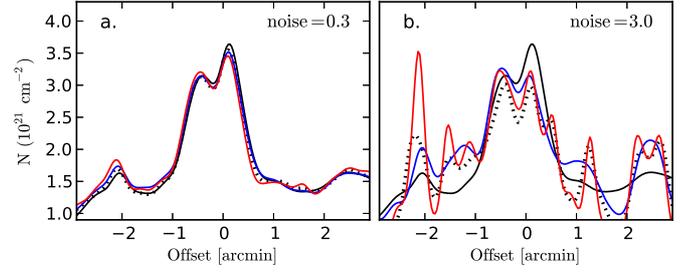}
\caption{
Column density profiles for a horizontal cut through the maps in
Fig.~\ref{fig:tff031_map}. The noise levels are given in the frames.
The lines show the true column density (black line), the column
density derived from 18$\arcsec$ resolution surface brightness data
(dotted line) and the results of Methods A and B (blue and red lines,
respectively).
}
\label{fig:tff031_profile}%
\end{figure}

\subsubsection{Estimates on large scales: Cloud I} \label{sect:extended}

We start by examining Cloud I. Because we are mostly interested in the
dense structures and because the bias is stronger at higher column
densities, we restrict the analysis to regions with true column
density $N_{\rm true} > 5\times10^{21}$\,cm$^{-2}$.
Figure~\ref{fig:tff031_colden} compares the results of Methods A and B
to the true column density, including the Pearson correlation
coefficients. On average, the methods give comparable results. For
lower observational noise, Method B gives higher correlation with the
true values. However, Method B is also more sensitive to the presence
of noise and, in the case of three times the default noise level, the
correlation coefficient is higher for Method A.

Figure~\ref{fig:tff031_map} compares the results in a small region centred at a
dense clump, i.e., concentrating on the highest column densities. The
calculations are carried out with the default observational noise. In the
figure, in the neighbourhood of the main peak, Method A now shows up to
$\sim$50\% greater errors than Method B. This appears to be a direct consequence
of the low resolution of the temperature information used by Method A. The
figure also suggests that a large fraction of the errors in Method B map result
from temperature variations that increase with the line-of-sight column
density. Therefore, the availability of surface brightness data at the
18.3$\arcsec$ resolution would not result in lower errors.
Figure~\ref{fig:tff031_profile} shows column density profiles for the same
region, again comparing cases with 0.3 times at 3.0 times the default noise.

\begin{figure}
\centering
\includegraphics[width=8.7cm]{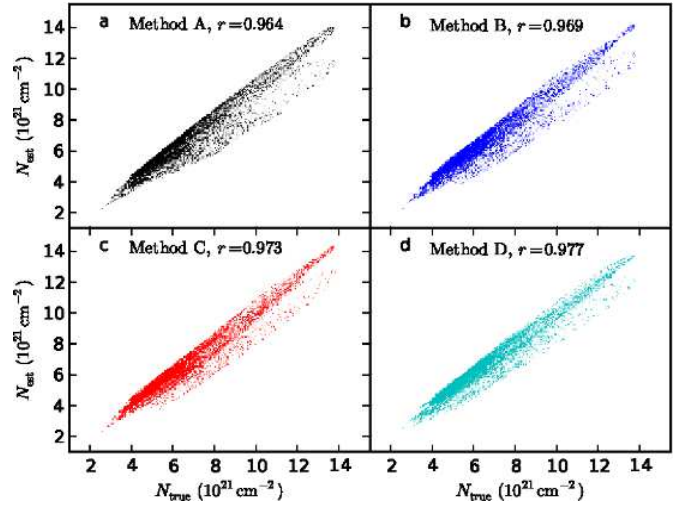}
\caption{
Correlations between column density estimates and the true column
density for Methods A, B, C, and D in the case of Cloud I and the
default noise level. The plots and the correlation coefficients listed
in the frames correspond to column density estimates that are
all convolved to a common resolution of 36.3$\arcsec$.
}
\label{fig:correlation_tff031}%
\end{figure}

We also examine Method C that is an optimised version of Method B, a
linear combination that results in the best correlation with the true
column densities at the 18.3$\arcsec$ resolution. Finally, Method
D is a similar linear combination of the $N(250 \rightarrow 500)$,
$N(350 \rightarrow 500)$, and $N(500)$ maps at the lower 36.3$\arcsec$
resolution. Figure~\ref{fig:correlation_tff031} shows the correlations
as scatter plots.

In these overall correlations, Method B is performing
consistently slightly better than Method A. With its optimised linear
coefficients, Method C produces still some improvement that is visible
as a smaller scatter in Fig.~\ref{fig:correlation_tff031}. Even more
interesting are the best coefficients that are obtained for the terms
$N(500)$, $\left[ N(350) - N(350\rightarrow 500) \right]$, $\left[
N(250) - N(250\rightarrow 350) \right]$ (see
Table~\ref{table:coeffs}). In Method B these are by construction all
equal to 1.0. However, in the case of Cloud I, both Method C and
Method D have partly negative coefficients. The correlation
coefficient with true values is highest for Method D, $r$=0.977,
partly also because of the lower resolution. 
The opposite signs and greater magnitude of the $N(250)$ and $N(350)$
coefficients mean that significant corrections are made
based on the small differences between the individual column
density estimates.

The results are partly encouraging, suggesting that it might be
possible to improve the accuracy of the column density estimates beyond
those obtained assuming a single modified black body. However, the
method should perform well for all clouds, not only for the one for
which it is originally tuned. 
We repeated the previous analysis also using $\beta=1.7$ instead of
$\beta=2.0$. This changes the coefficients of Methods C and D by some
tens of percent but the correlations with the true column density are
only slightly lower (less than 0.01). The main effect of a wrong 
spectral index value is, of course, bias in the absolute values of the
column density estimates.

\subsubsection{Estimates for cores: Cloud II} \label{sect:cores}

In this section, we concentrate on the densest structures of Cloud II (the
cores) and their environment.  We look first at all data between the column
densities of $5\times10^{21}$\,cm$^{-2}$ and $20\times10^{21}$\,cm$^{-2}$, the
selection being made using the true column densities at a resolution of
18.3$\arcsec$. The lower limit is the same as above while the upper limit is
$\sim$25\% higher than the maximum column density of Cloud I. For these data
all Methods A--D give a correlation coefficient $r \sim 0.997$ for the
comparison with the true column density. This is probably due to the fact that
in Cloud II these column densities still correspond to extended medium without
strong dust temperature variations. In this situation the coefficients of
Methods C and D are not very well defined. 

In the column density range $20\times10^{21}$\,cm$^{-2}$ and
$100\times10^{21}$\,cm$^{-2}$, small differences again appear between
the methods. The correlation coefficients are $r=0.972$ for Method A,
$r=0.978$ for both Method B and Method C, and $r=0.987$ for Method D.
Unfortunately, for Method D, the coefficients found in Cloud I and
Cloud II or between the column density intervals of Cloud II are not
very similar (see Table~\ref{table:coeffs}). For Method C, the
variation of the parameter values is smaller. Nevertheless, there is
no single set of coefficients that could be used to improve the
accuracy of the column density estimates for any cloud.

\begin{table*}
\caption{Linear coefficients of Methods C and D. In Method C the
column density estimate is 
$ p_1 \times N(500) + 
  p_2 \times \left[ N(350) - N(350\rightarrow 500) \right] + 
  p_3 \times \left[ N(250) - N(250\rightarrow 350) \right]$, 
in Method D the estimate is 
  $p_1 \times N(250 \rightarrow 500) + p_2 \times N(350 \rightarrow
  500) + p_3 \times N(500)$.
The second column specifies the selection of the analysed region, based on column
density or the area around the selected cores.
}
\label{table:coeffs}
\begin{tabular}{llcccccc}
Model cloud  &  $N({\rm H})$  &  \multicolumn{3}{c}{Method C}
&  \multicolumn{3}{c}{Method D}  \\
  &  ($10^{21}$\,cm$^{-2}$) & $p_1$ & $p_2$ & $p_3$   & $p_1$ & $p_2$  & $p_3$  \\
\hline
I       &   $>$5    &  +1.00  & +2.61 & -0.37  &   -6.80 & +7.20 & +0.88  \\
II      &   5--20   &  +0.99  & +0.28 & -0.05  &   -0.01 & -0.32 & +1.32  \\
II      &   20--100 &  +1.03  & +2.56 & -0.76  &   -1.75 & -0.23 & +3.09  \\
II      & cores     &  +1.08   &  +6.54 & -2.48  &   -4.47 & -0.12 & +5.83  \\
II w. sources 
        & cores     &  +1.09   &  +5.22 & -0.82  &   -4.46 &  1.60 & +4.18  \\
\hline
\end{tabular}
\end{table*}

On large scales, the linear combinations of $N(250)$, $N(350)$, and
$N(500)$ did produce some improvement in the accuracy but not with
universal coefficients. The usual column density estimates should be
more biased at the locations of the dense cores where the
line-of-sight temperature variations are the greatest. The small scale
fidelity of the column density maps is crucial for the interpretation
of the core properties. Therefore, we examine separately the
neighbourhood of the cores in Cloud II. These are the same
gravitationally bound regions as discussed in \cite{Malinen2011}. We
study pixels that fall within 2$\arcmin$ radius of the centre of each
core.

\begin{figure}
\centering
\includegraphics[width=8.7cm]{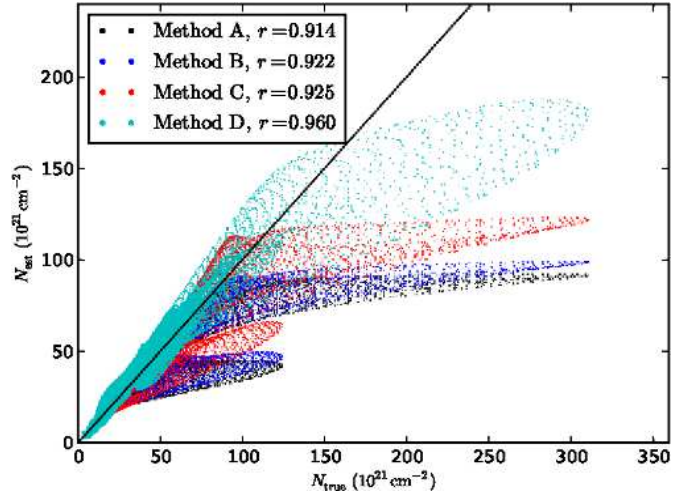}
\caption{
Correlations between column density estimates and the true column
density. The figure compares the results of Methods A, B, C, and D
using data around the cores of Cloud II. The estimates have been all
convolved to a common resolution of 36.3$\arcsec$ and scaled so that
the median values fall on the correct relation $N=N_{\rm true}$ (solid
line).
}
\label{fig:collins_BG_cores}%
\end{figure}

Figure~\ref{fig:collins_BG_cores} compares the results at the 36.3$\arcsec$
resolution. All methods tend to underestimate the true column density,
especially towards the column density peaks. This produces the strong
flattening of column densities above $\sim$10$^{22}$\,cm$^{-2}$ and
the greatest
errors are close to a factor of three. The order of accuracy of Methods A--D is
as in the case of the large scale correlations but the differences are more
pronounced. At the highest column densities, the estimates given by Method D
are twice the values of Methods A and B. 
%%%
Nevertheless, even Method D underestimates the true column
density by up to $\sim$50\%.
For this set of data, the correlation
coefficients are 0.914, 0.921, 0.925, 0.964 for Methods A--D,
respectively.

\begin{figure}
\centering
\includegraphics[width=8.7cm]{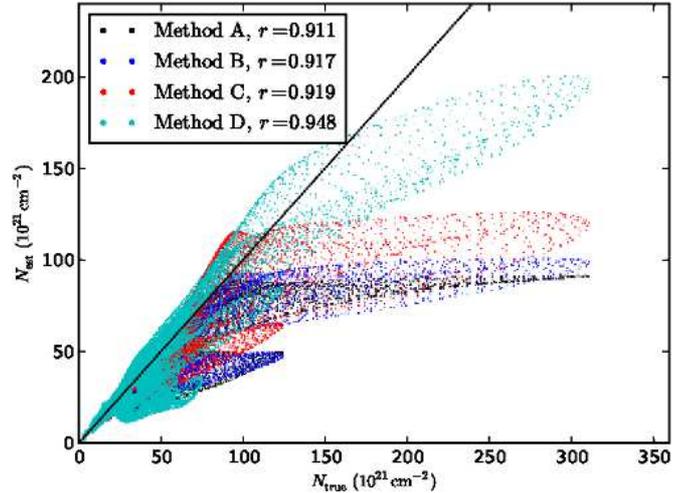}
\caption{
As Fig.~\ref{fig:collins_BG_cores} but including internal heating
of the cores.
}
\label{fig:collins_SOU_cores}
\end{figure}

\cite{Malinen2011} noted that when the cores are extremely dense,
internal heating will improve the accuracy of the column density
estimates. However, the effect was small for the Cloud II (see their
Fig. 11) and depends on how large regions around the cores are
examined. We plot in Fig.~\ref{fig:collins_SOU_cores} similar
relation but for all pixels within the 2$\arcmin$ radius. The figure
is thus similar to Fig.~\ref{fig:collins_BG_cores} but shows the
situation after the addition of the internal heating sources. 

The sources only have a little effect on
the accuracy of the column density estimates when this is calculated
at the resolution of the 500\,$\mu$m data, the FWHM corresponding to
$\sim$20 pixels. The bias shown by Methods A--C is very similar to
Fig.~\ref{fig:collins_BG_cores} but the correlation coefficients are
slightly lower. At the highest column densities, Method D manages to
bring the column density estimates up by $\sim$10\% but below
100$\times 10^{21}$\,cm$^{-2}$ the results show a greater
scatter. The coefficients optimised for the cores differ from those
derived on large scale but, on the other hand, the effect of the
internal heating sources remains marginal (see
Table~\ref{table:coeffs}).

\begin{figure}
\centering
\includegraphics[width=8.7cm]{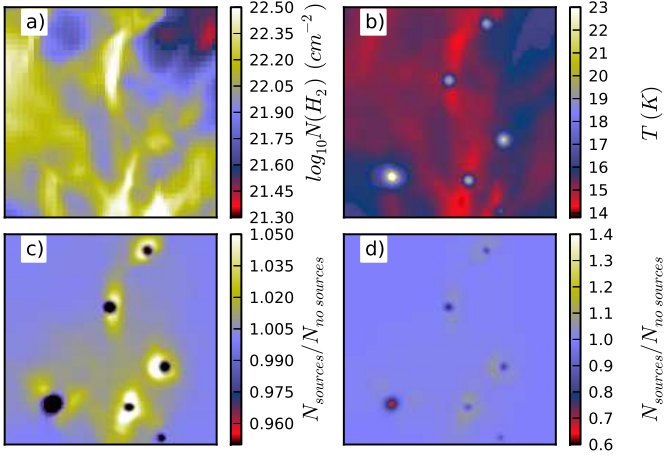}
\caption{
Column density errors for Cloud II with internal sources. The upper
frames show the true column density and the estimated colour
temperature. The lower frames show, with different colour scales, the
ratio of the column density estimates with and without internal
sources. The maps show a $20\arcmin \times 20\arcmin$ area of the
full model. The estimates are calculated using 160--500\,$\mu$m
surface brightness maps with 18.3$\arcsec$ resolution.
}
\label{fig:BG_SOU}
\end{figure}

Figure~\ref{fig:BG_SOU} shows a 2.9\,pc$\times$2.9\,pc piece of Cloud
II, looking at the column density estimates obtained from surface
brightness data at a resolution of 18.3$\arcsec$. The area includes
some filamentary structures and six sources that raise the colour
temperature locally up to $\sim$25\,K.  With the sources, the column
density estimates of the main filament are higher by up to $\sim$10\%.
This might be related to a higher average dust temperature that
decreases the bias associated with line-of-sight temperature
variations. At the very location of the sources the column density
estimates are 10--30\% lower. If the column density maps are convolved
to a 2$\arcmin$ resolution, the internal sources are seen to increase
(i.e., improve) the estimates of the average column density of the
cores. The net effect is negative only for a few of the strongest
sources. This is the case for the source in the lower left corner of
Fig.~\ref{fig:BG_SOU}) where, at the 2$\arcmin$ resolution, the
estimate of the average column density of the region is still by
$\sim$7\% lower because of the presence of the radiation source. At
the 18.3$\arcsec$ resolution, the effect is 35\% towards the source
but goes to zero already at the distance of $\sim$20$\arcsec$.

\begin{figure}
\centering
\includegraphics[width=8.7cm]{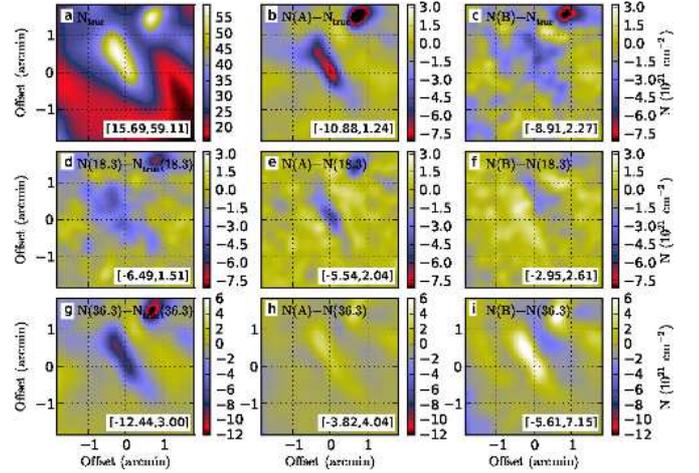}
\caption{
Comparison of column density estimates in a 0.5\,pc$\times$0.5\,pc
area in Cloud II. Frame $a$ shows the true column density and frames
$d$ and $g$ the errors for column density maps derived with all data
at a resolution of 18.3$\arcsec$ or 36.3$\arcsec$. The second and the
third columns show the difference between Method A and Method B
relative to the true column density (first row) and the estimates
obtained with all surface brightness data either at 18.3$\arcsec$
(second row) or 36.3$\arcsec$ resolution (third row). The numbers at the bottom of each
frame show the range of values within the map.
}
\label{fig:collins_map_BG}%
\end{figure}

\begin{figure}
\centering
\includegraphics[width=8.7cm]{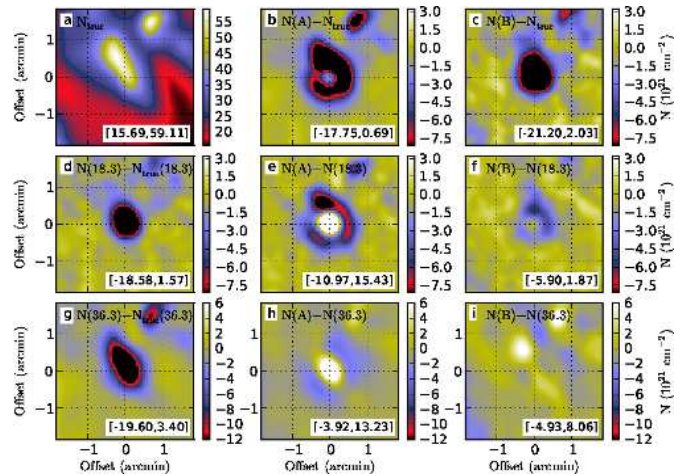}
\caption{
As Fig.~\ref{fig:collins_map_BG} but showing the same 
0.5\,pc$\times$0.5\,pc area of Cloud II with an added internal heating
source in the middle of the field. The 33.7 solar luminosity source
raises the dust colour temperature locally to $\sim$27\,K. 
}
\label{fig:collins_map_SOU}%
\end{figure}

\begin{figure}
\centering
\includegraphics[width=8.7cm]{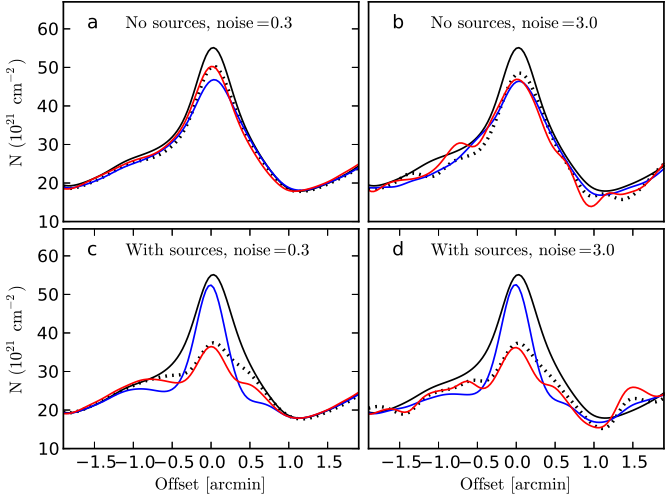}
\caption{
Column density profiles for a horizontal cut through the maps 
Figs.~\ref{fig:collins_map_BG}-\ref{fig:collins_map_SOU}.  Shown are
the true column density (solid black line; uppermost lines, convolved
to 18.3$\arcsec$ resolution), the estimate derived from 18.3$\arcsec$
data at all wavelengths (black dotted line), Method A (blue solid
line), and Method B (red solid line). The lower frames correspond to
the case with an internal heating source. The noise is 0.3 times
(frames $a$ and $c$) or 3.0 times (frames $b$ and $d$) times the
default value.
}
\label{fig:collins_profile}%
\end{figure}

Figures~\ref{fig:collins_map_BG}-\ref{fig:collins_profile} take a closer look
at a $\sim$0.5$\times$0.5\,pc area in Cloud II. The predictions of Method A and
Method B are compared in Fig.~\ref{fig:collins_map_BG} with the results
obtained directly using surface brightness data at a resolution of
18.3$\arcsec$ or 36.3$\arcsec$. For the main clump in the field
(containing a gravitationally bound core), the maximum error of Method A is
about twice as large as for Method B. The results of Method B are very similar
to the map derived from 18.3$\arcsec$ data. Both contain errors caused by the
line-of-sight temperature variations. The maximum error is found at the
position of a dense core in the upper right hand part of the figure where the
error of Method B is $\sim$37\% greater than for the 18.3$\arcsec$ data.
However, this error is still less than 20\% when compared to the true column
density. When the estimation is based on data at a resolution of 36.3$\arcsec$,
the column density map appears smooth, because of the larger beam and because
of the higher signal-to-noise ratio. In spite of the lower resolution, the
maximum errors are greater than for either Method A or Method B at twice
as high resolution.

Figure~\ref{fig:collins_map_SOU} shows the same area after the addition of
an internal heating source. Because of stronger temperature variations, the
column density errors are $\sim$30\%, the 36.3$\arcsec$ surface brightness data
resulting in only slightly greater errors than the 18.3$\arcsec$ data. 
Compared to Method B, the errors of Method A are smaller at lower column
densities (because of lower sensitivity to noise), greater in a ring around
the main clump, and again smaller at the location of the heating source where
the error is below 10\%. Figure~\ref{fig:collins_profile} shows the column
density profiles for the main clump. For the quiescent core, the Method B gives
a profile that is almost identical to that of high resolution data, this still
slightly underestimating the true column density. In Method A, the low
resolution of the temperature information results in too low column densities
for the quiescent core but in the case of internally heated core this compensates
the effect of temperature variations. In the case of internal heating, none of
the methods is able to recover the actual column density profile of the clump.
The result of Method B also is clearly impacted when the observational noise is
increased.

\begin{figure}
\centering
\includegraphics[width=8.7cm]{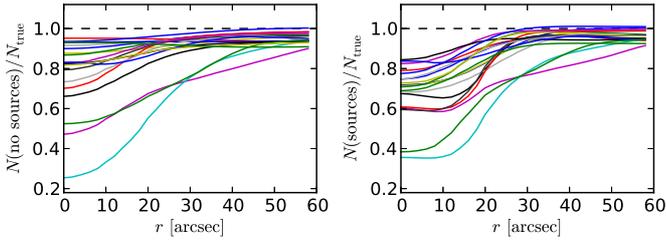}
\caption{
The ratio of the column density estimates and the true values in
the 19 cores of Cloud II with column densities $N>5\times
10^{22}$\,cm$^{-2}$. The results are shown separately for the case
without internal heating (left frame) and with one radiation source
inside each core (right frame). The estimates are calculated using
18.3$\arcsec$ resolution surface brightness data at all wavelengths
160--500\,$\mu$m. 
}
\label{fig:radial}
\end{figure}

\subsection{Non-linear correction to column density estimates}
\label{sect:nonlinear}

In this section we examine non-linear combinations of the $N(250)$,
$N(350)$, and $N(500)$ values. The goal is to find a method that would
better trace the relative column density variations around the dense
cores.

Figure~\ref{fig:radial} shows the radial relative error of the column
density estimates for all cores with peak column density exceeding
$5\times 10^{22}$\,cm$^{-2}$. The values are calculated using data at
18.3$\arcsec$ resolution, i.e., the figure shows only the effect of
the line-of-sight temperature variations on the estimates. The column
density is underestimated up to 70\% or, in the case of internal
heating, up to $\sim$60\%. For most cores the errors are below 30\%.

\begin{figure}
\centering
\includegraphics[width=8.7cm]{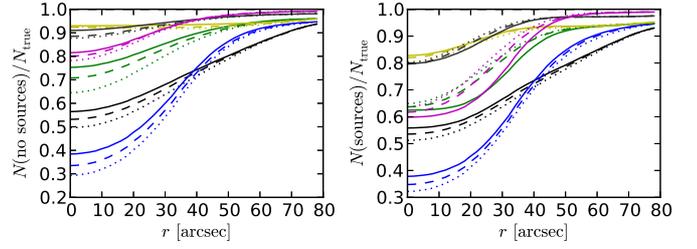}
\caption{
The ratio of the column density estimates and the true values in
a sample of six Cloud II cores with $N>5\times 10^{22}$\,cm$^{-2}$.
The dotted, dashed, and solid lines correspond to $N(250)$, $N(350)$,
and $N(500)$, respectively, all calculated at the 18.3$\arcsec$
resolution. The results are shown for cores without (left frame) and
with internal heating sources (right frame). 
}
\label{fig:radial_NL}
\end{figure}

Figure~\ref{fig:radial_NL} shows some examples of the radial profiles
for the estimates $N(250)$, $N(350)$, and $N(500)$ separately. Without
internal heating, the behaviour is very predictable with $N(250) <
N(350) < N(500)$. Therefore, the difference between the estimates can
be used to correct for the under-estimation of the column density.
However, Fig.~\ref{fig:radial_NL} also indicates that the fraction by
which column density is under-estimated is not completely systematical
relative to, for example, the ratio $N(500)/N(250)$.  With
internal heating the situation becomes more complicated. Within the
sample of six cores included in Fig.~\ref{fig:radial_NL} (right
frame), in two cases $N(250)$ is larger than $N(500)$. Such
differences are to be expected, because the column density estimates
depend in a complex way on the source luminosity and the geometry and
optical depth of the cores.

\begin{figure}
\centering
\includegraphics[width=8.7cm]{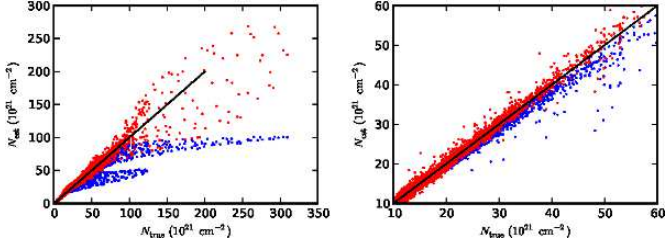}
\caption{
Correlations between the true column density and the estimates $N(500)$
(blue points) and $N_{\rm NL}$ (red points, see text). The data
consists of the 2$\arcmin$ radius environments of the cores in Cloud
II. The resolution is 36.3$\arcsec$. The right hand frame shows a
zoom-in to small column densities.
}
\label{fig:collins_NL}
\end{figure}

As the simplest extension of Method D,  we attempted a non-linear
correction using the formula $N_{\rm NL} = p_1 N(250 \rightarrow 500)+
p_2 N(350 \rightarrow 500) + p_3 N(500) + p_4 \left[ N(500)-N(250
\rightarrow 500) \right]^2$. For the 2$\arcmin$ neighbourhoods of the
gravitationally bound cores of Cloud II, the least-squares
solution
leads to the parameter values
%%%%
$p_1=-2.93$, $p_2=-0.042$, $p_3=4.10$, and $p_4=0.62$. 
%%%%
As suggested by Fig.~\ref{fig:radial_NL}, the greater the difference in
$N(500)-N(250)$, the greater the upward correction of the column density estimates.
The correction (see Fig.~\ref{fig:collins_NL}) increases the correlation
coefficient between the true column density and the estimates from 0.924 for
$N(500)$ to 0.976 for $N_{\rm NL}$. The coefficients are largely determined by the
two cores with the highest column densities. However, also in the range $N_{\rm
true}=(10-60) \times 10^{21}$\,cm$^{-2}$ the bias appears to be somewhat smaller
and the correlation coefficient has risen, although only very marginally (from
0.992 to 0.993). Of course, the quadratic term in the formula of $N_{\rm NL}$
does not take the sign of the $N(500)-N(250)$ difference into account.
In
practice, the results would not change if the difference were replaced with
max$\{0.0, N(500)-N(250)\}$.
The use of an additional term $\left[ N(500)-N(350) \right]^2$ does not
bring any further improvement.

When the same model is fitted to the data with internal sources,
the parameters are
%%%%
$p_1=-3.03$, $p_2=1.42$, $p_3=2.79$, and $p_4=0.48$.
%%%%
The parameters $p_1$, $p_3$, and $p_4$ are roughly similar as in the case
without sources,  but $p_2$ has increased to a value of 1.42. With
these parameters, the correlation with the true column density increases from
0.920 for $N(500)$ to 0.982 for $N_{\rm NL}$. If we use directly the
coefficients derived from the model without internal sources, the correlation
coefficient for $N_{\rm NL}$ becomes 0.977. This is still a clear improvement
over $N(500)$ but, with these coefficients, $N_{\rm NL}$ overestimates the
column densities beyond $\sim150 \times 10^{21}$\,cm$^{-2}$, the error
increasing to $\sim$30\% for the highest column densities. However, for these
lines-of-sight $N(500)$ underestimates the true values by a factor of three.
This shows that the estimated non-linear correction could be useful in
general, not only in the cloud where its coefficients were derived.

\section{Radiative transfer modelling} \label{sect:modelling}

The errors in the predictions of the previous methods are largely
related to the temperature variations within the sources and the
way the variations are reflected in observations at different
wavelengths. By
constructing three-dimensional models of the sources, one should be
able to account for these effects. In this section, we
examine how well this works in practice, especially in the case of
high column densities. Our synthetic observations are themselves
based on numerical simulations. In this section, we only use
the resulting surface brightness data, not the information about the 
structure of the sources or the radiation field seen by the
individual clumps.

\subsection{The modelling procedures}  \label{sect:procedures}

\begin{figure}
\centering
\includegraphics[width=8.7cm]{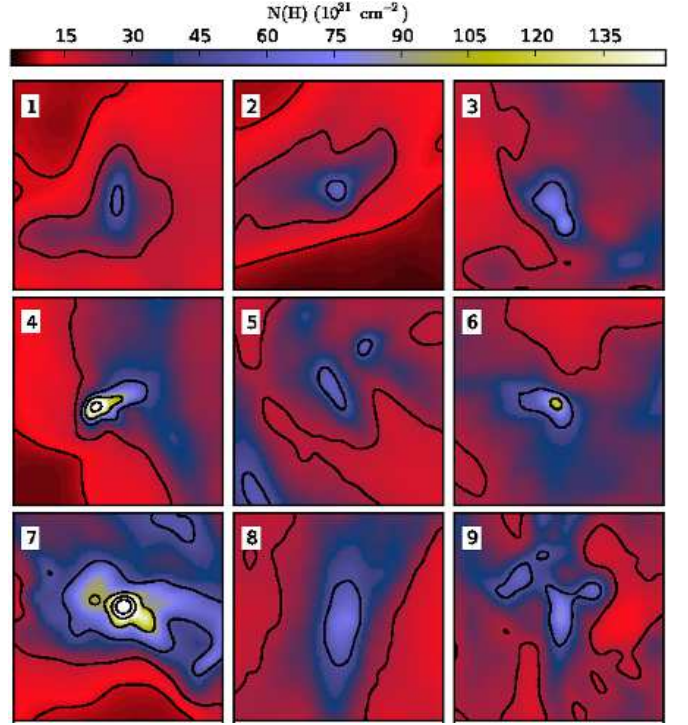}
\caption{
Column density maps of the nine fields (cores) selected for radiative transfer
modelling. The contours are at 10.0, 20.0, 50.0, 100.0, 200.0, and 400.0 times
$10^{21}$\,cm$^{-2}$.
}
\label{fig:9N}%
\end{figure}

We carry out radiative transfer modelling of the nine cores of
Cloud II with the highest column densities. The modelling is done
purely on the basis of the ``observed'' surface brightness maps,
without using any information on the three-dimensional structure of
the cloud.
Each core separately is described with model of 81$^3$ 4$\arcsec$
cells, the cell size corresponding to 0.0098\,pc. Each model thus
covers a projected area of $5.3\arcmin \times 5.3\arcmin$ or
0.79\,pc$\times$0.79\,pc. The column density maps are shown in
Fig.~\ref{fig:9N} with values obtained from the Cloud II density
cube. Note that the term ``observation'' refers to
the synthetic surface brightness maps obtained from Cloud II and the
term ``model'' refers to the 81$^3$ cell models constructed for the
regions around the selected cores.

Along the line-of-sight, the full extent of Cloud II is 10\,pc, much
greater than the size of the $81^3$ cell models. Thus, the structures
that are visible within these $5.3\arcmin \times 5.3\arcmin$ maps may
be physically connected or may be quite distant from each other. The
lack of knowledge on the line-of-sight structure is one of the main
sources of uncertainty in the modelling. To analyse the observations
of Cloud II, we subtract from the surface brightness maps the local
background that is defined with the position of minimum $N(500)$
value within the $5.3\arcmin \times 5.3\arcmin$ area. This eliminates
the necessity of modelling the extended emission. 
%%%
In some fields surface brightness gradients result in large regions
with negative residual signal. However, this has little effect on the
actual cores that are much above the background.
%%%
The line-of-sight density distribution is assumed to be Gaussian, with
the FWHM set to 20\% of the box size. The value is not tuned
separately for each core but it is roughly consistent with the average
core sizes in the plane of the sky. In the plane of the sky, the
models are optimised by comparing the observed surface brightness
maps and the corresponding maps produced by the models. The comparison
is restricted to inner $2\arcmin \times 2\arcmin$ area because the
boundaries are affected by edge effects (because of the flat surfaces
subjected to the full external field).

The modelled cores are embedded deep within Cloud II that has an average visual
extinction of $A_{V}\sim$6.6$^{\rm mag}$. This means that the radiation
field is different for each core. In the calculations, this is taken into
account by illuminating the model cloud by an attenuated radiation field. The
unattenuated radiation field corresponds to three times the \citet{Mathis1983}
model of the interstellar radiation field (ISRF), i.e., it is clearly higher
than the actual field in the original simulations. The attenuation is
parametrised by the visual extinction $A_{V}^{\rm ext}$ and is calculated
using the extinction curve of the selected dust model. The attenuation is
assumed to take place outside the model volume, by some external diffuse cloud
component that would correspond to the diffuse component already subtracted
from observations.

We carry out radiative transfer calculations to produce synthetic 81$\times$81
pixel maps of surface brightness that are compared with the ``observations'' of
Cloud II. The model column densities are adjusted according to the ratio of the
observed and the modelled 250\,$\mu$m values. This means that the density in
each cell corresponding to a given map pixel is multiplied by the same number
that depends on whether the current model is overestimating or underestimating
the observed surface brightness. The procedure is iterated until the
250\,$\mu$m errors are below 1\% for the innermost $2\arcmin \times 2\arcmin$
area.

\subsubsection{Basic models} \label{sect:basic}

If we do not use any other spectral information, the column density estimates
depend on the assumed intensity of the heating radiation. If the same value
were used for all cores, the column densities would often be wrong by a factor
of two, a result significantly worse than either of Methods A or B
(comparing the estimates to the {\em true} column density in Cloud II).
Therefore, we adjust the attenuation of the external field until the observed
and modelled ratios of 160\,$\mu$m and 500\,$\mu$m surface brightness agree.
The ratios are measured as the average within a 30$\arcsec$ radius of the core.
Thus, both the intensity and shape of the emission spectral energy distribution
(SED) should be correct at the location of the column density peak. Both
attenuation and column densities (one value per map pixel) are adjusted
iteratively until the relative errors are below 1\%.

\begin{figure}
\centering
\includegraphics[width=8.7cm]{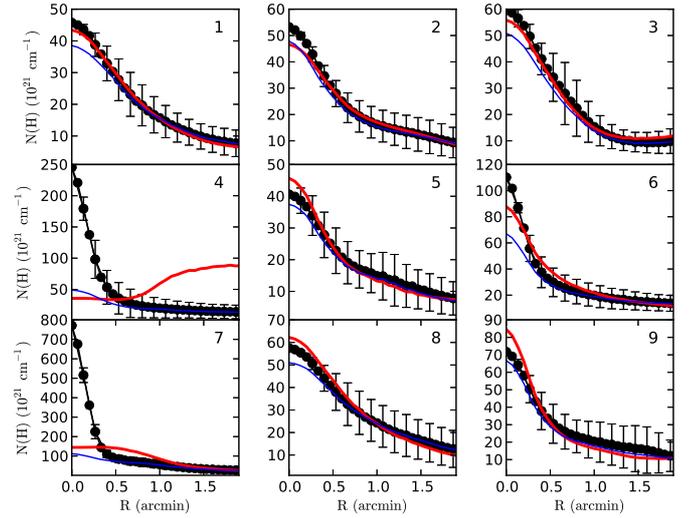}
\caption{
Azimuthally-averaged column density profiles of the selected nine cores. The
black symbols and error bars show the true column density and the variation in
4$\arcsec$ wide annuli. The blue solid curve is the estimate from Method B and
the red solid curve the profile obtained from the constructed radiative
transfer model.
}
\label{fig:9B}
\end{figure}

Figure~\ref{fig:9B} compares the column densities of the optimised
models to the true column densities and to the estimates from Method
B. Although the cores are not circularly-symmetric, we plot the
azimuthally averaged column densities as the function of distance from
the centre of the selected core. The error bars on the true
column density correspond to the azimuthal variation. For cores 1--3,
the modelling recovers the radial column density profiles quite
accurately. For core 6, the fit is worse but still better than for
Method B that underestimates the true values by a significant
fraction. For core 8, the Method B and modelling are equally
close to the truth, while in cores 5 and 9 the modelling overestimates
the peak. For the sources with the highest column densities, cores number 4 and
7, the peak is missing for both methods. This is not surprising, as the
very compact central object is almost invisible still at the
wavelength of 250\,$\mu$m.  

If the external field is raised to five times the \cite{Mathis1983} values, the
central column densities decrease by 5--10\%. The higher level of the external
field is compensated by a greater value of $A_{V}^{\rm ext}$, which means
that energy is absorbed mainly at longer wavelengths. This reduces temperature
gradients, also reducing the estimated column density contrasts of the cores by
a similar factor of $\sim$10\%.

\subsubsection{Asymmetric illumination} \label{sect:asymmetric}

In the field number 4 the results of the modelling are completely wrong at
greater distances (Fig.\ref{fig:9B}). This is caused by the presence of a strong
temperature gradient across the field. The SED is correct at the centre of the
field but on one side of the map the same assumption of the external field
strength is not enough to produce the observed 250\,$\mu$m intensity.  The
column density increases to the point where the surface brightness saturates,
leaving a spot where the 250\,$\mu$m surface brightness is underestimated in spite
of the model column density being far above the correct value.
%% @@
Fortunately, the problem is plainly visible in the surface brightness
maps, i.e., is apparent for the observer. In the problem area, the
model underestimates the 160\,$\mu$m surface brightness by more than
50\% and overestimates the 500\,$\mu$m intensity by $\sim$150\%.

Temperature gradients are seen in a few fields and these can affect also the
central column density because of the mutual shadowing of the regions. To make
a first order correction for gradients in the plane of the sky, we added to
each model an anisotropic radiation source that covers a circular area of the
sky with an opening angle of 45 degrees. The centre of this sky area is in a
direction perpendicular to the line of sight and at a position angle where,
based on the previous results, the source will help to remove the residual
colour gradients. The spectrum of the anisotropic component corresponds to the
normal ISRF attenuated by $A_{V}=2.0^{\rm mag}$ and its intensity is scaled
to obtain a solution where, for data within the innermost one arcmin radius,
the quantity $\Delta S(160\mu{\rm m})-\Delta S(500\mu{\rm m})$ is no longer
correlated with the distance along the original gradient direction. In the
formula, $\Delta S$  stands for the difference of the observed and modelled
surface brightness values.

\begin{figure}
\centering
\includegraphics[width=8.7cm]{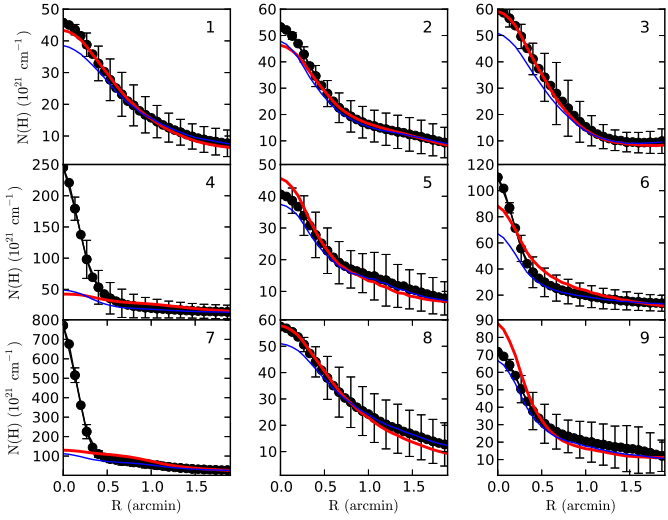}
\caption{
Azimuthally-averaged column density profiles of the selected nine cores. As
Fig.~\ref{fig:9B} but including in the modelling anisotropic radiation field.
}
\label{fig:9C}
\end{figure}

\begin{figure}
\centering
\includegraphics[width=8.7cm]{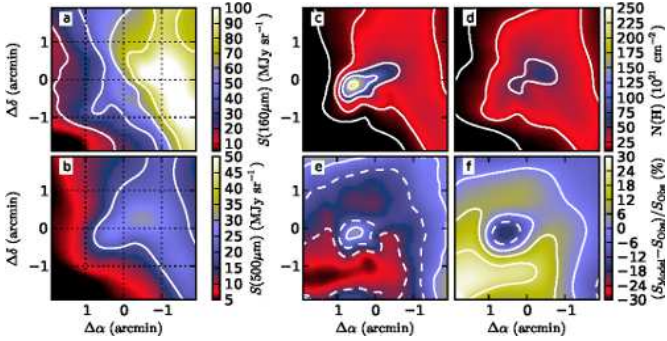}
\caption{
Core 4 modelled with anisotropic radiation field. Frames $a$ and $b$ are the
160\,$\mu$m and 500\,$\mu$m surface brightness maps from Cloud II (the
``observations''). The core is visible in the map of the true column density
(frame $c$) but is missed by the constructed model (frame $d$). Frames $e$ and
$f$ show the errors of the model predictions $S_{\rm Model}$ relative
to the observed surface brightness $S_{\rm Obs}$ at the wavelengths of
160\,$\mu$m and 500\,$\mu$m, respectively.
}
\label{fig:C16}
\end{figure}

The effect of asymmetric illumination is in most cores almost unnoticeable on
the column density estimates.  In core 3, the central column density
increases by $\sim$5\% while in core 8 it decreases by the same amount. Clear
effects are visible only in the case of the cores 4 and 7, the ones with the
highest column density (see Fig.~\ref{fig:9C}). In field 4, the model is now
much closer to the correct column density values outside the central core.
However, the constructed models still miss the high column density peaks of
both fields 4 and 7. Figure~\ref{fig:C16} shows this for core 4. The densest
core is invisible in the 160\,$\mu$m map, not visible as a separate peak in
the 500\,$\mu$m surface brightness, and also missed by the model. The model
was tuned so that the SED (250\,$\mu$m surface brightness and the
160\,$\mu$m/500\,$\mu$m colour) on the average match the observations over an
one arcmin circle.  At the very centre, however, the residual errors are
$\sim$10\% of the surface brightness and rise above 30\% elsewhere in the
field. The significance of these residuals suggests that further improvements
in the models should be possible.

\subsubsection{Varying the line-of-sight mass distribution}
\label{sect:LOS}

The main problem for the modelling is that the cores are embedded in an optically
thick cloud whose line-of-sight extent is two orders of magnitude longer than the
typical core size. Any structure seen in a map can be a compact object also in
three-dimensions, it may be elongated along the line-of-sight or may even consist of
several unconnected structures within the 10\,pc distance through the model cloud.
This impacts the dust temperatures and, consequently, the column density estimates.

The previous models consisted of a smooth density distribution with a
single FWHM for its line-of-sight extent. We can try to take some of
the variations into account by modifying the FWHM values for each
line-of-sight separately.  A small FWHM value would corresponds to a
compact and cold region, a greater FWHM to more diffuse region with a
higher average temperature. We do not modify the FWHM at the core
location where the data are already used to adjust the strength of the
external radiation field. For any other line-of-sight, if the model
predicts a too cold spectrum (low ratio of 160\,$\mu$m and 500\,$\mu$m
intensities), the FWHM is increased (and vice versa). The final effect
is complicated by shielding between the different regions.

\begin{figure}
\centering
\includegraphics[width=8.7cm]{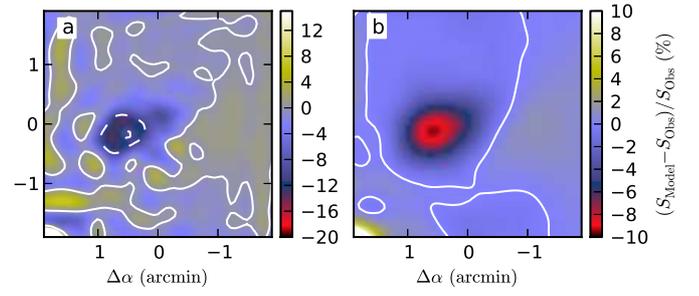}
\caption{
Core 4 modelled with anisotropic radiation field and varying line-of-sight
width of the density distribution. The maps show the relative error of the
model predictions at 160\,$\mu$m (frame $a$) and at 500\,$\mu$m (frame $b$).
The contours are at intervals of 10\%.
}
\label{fig:D16}
\end{figure}

This further modification helps to bring down the errors outside the densest core,
e.g., in the field number 4 (see Fig.~\ref{fig:D16}). The surface brightness
residuals at 160\,$\mu$m and 500\,$\mu$m are mostly below 10\% but raise up to 20\%
at the location of the column density peak. The column density estimate of the peak
is close to that of the previous models. The situation is similar in field 7 where
the peak value of $\sim120\times 10^{21}\,{\rm cm^2}$ remains far below the correct
number of $\sim 800\times 10^{21}\,{\rm cm^2}$.

Of the lower column density cores, the column density predictions were previously
most erroneous in field 9 where the peak value was overestimated by $\sim$35\%. The
adjustment of the line-of-sight extent of the clump has reduced this error only to
$\sim$28\%. For the other cores, the effects are smaller. This shows that even the
variation of the line-of-sight extent, as implemented, is not able to reproduce the
complexity of these regions.

\subsection{Ellipsoidal model}  \label{sect:ellipsoid}

\begin{figure}
\centering
\includegraphics[width=8.7cm]{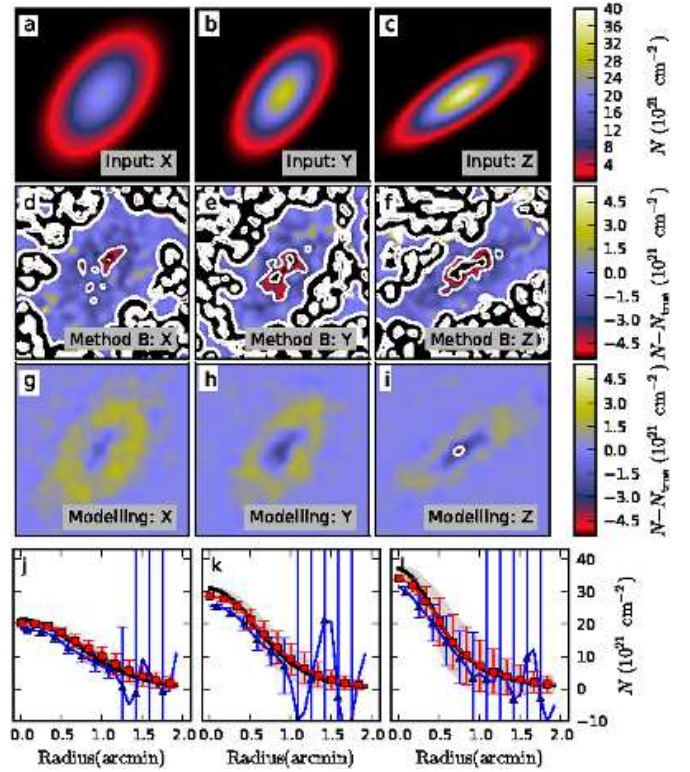}
\caption{
Results for the ellipsoid model. The frames $a$--$c$ show the true
column densities towards three directions. The errors in the column
density estimates of Method B and of the 3D modelling are shown in
frames $d$--$f$ and $g$--$i$, respectively, with contours drawn at
$-3\times 10^{21}$cm$^{-2}$ and $-6\times 10^{21}$cm$^{-2}$.
The bottom frames show the radial column density profiles. The true
column density is shown with a black solid line, the grey area
indicating the 1 $\sigma$ variation in the averaged rings. The
triangles correspond to Method B and the square symbols to the 3D
modelling. The error bars indicate the corresponding 1 $\sigma$
variation in the averaged rings.
}
\label{fig:ellipsoid}
\end{figure}

For the final test, we return to a simple geometry and isotropic
illumination. The cloud densities are generated as a three-dimensional
ellipsoid with Gaussian density profiles and a ratio of 1:2:4 between
the FWHM values along the main axes. The density field is sampled on a
81$^3$ Cartesian grid, first applying three 45 degree rotations to the
density distribution. The resulting cloud has peak column densities of
21.5$\times 10^{21}$cm$^{-2}$, 31.2$\times 10^{21}$cm$^{-2}$, and
37.5$\times 10^{21}$cm$^{-2}$ towards the three main axes (see
Fig.~\ref{fig:ellipsoid}). We calculate synthetic surface brightness
maps with the default noise and ISRF. Based on these data, the
column densities were then estimated with 3D modelling similar to that
of Sect.~\ref{sect:basic}. Thus, in the modelling only the column
densities and the external isotropic field were adjusted. The results
are compared only with the results of Method B because, as seen in
Sect.~\ref{sect:results}, the differences in the accuracy of Methods
A--D are not very significant.

When the 3D modelling was done with the correct level of the external fields
but keeping its attenuation as a free parameter, the column densities were
recovered with an accuracy of a couple of percent. The cloud has a density
distribution that is consistent with the assumption of a Gaussian line-of-sight
density distribution used in the modelling. However, the width of the
distribution is not the same as in the modelling and, furthermore, varies by a
factor of four depending on the viewing direction. This suggests that in this
case the results are not very sensitive to the uncertainty of the line-of-sight
extent.

The results of Method B and the 3D modelling are compared further in
Fig.~\ref{fig:ellipsoid}. In this case, the modelling is done with an external
field that is twice as strong as the actual field used to produce the synthetic
observations. Therefore, the field needs to be adjusted by introducing
significant external attenuation. This correction is not exact because the
attenuation changes not only the level but also the SED of the incoming
radiation. After removing the shortest wavelengths, the remaining radiation
penetrates deeper, making the cloud more isothermal. The resulting errors are
visible in Fig.~\ref{fig:ellipsoid}g-i where the column density is
overestimated in the outer parts of the cloud and correspondingly
underestimated at the centre. The errors rise over $\sim$4\% only at the centre
and only when the cloud is viewed from the direction with the highest column
density. The incorrect assumption of the ISRF spectrum is also visible in the
residuals at 160\,$\mu$m and 500\,$\mu$m. Although the average colours are
adjusted to be correct, at the cloud centre 160\,$\mu$m intensity is
overestimated by more than $\sim$5\% while the 500\,$\mu$m intensity is
underestimated by a couple of percent. This information could be used to
further improve the accuracy of the model. For Method B the errors are
stronger,
column density being underestimated by up to $\sim$10\% percent. The relative
bias is better visible in the radial profiles at the bottom of
Fig.~\ref{fig:ellipsoid}.

The sensitivity to noise is another important point. Unlike in the Cloud I and
Cloud II, the observations of the outer parts of the clump are now dominated by
noise. This affects Method B results already at $\sim 4 \times
10^{21}$cm$^{-2}$, mainly via the $N(250)$ estimates. The modelling results in
much lower statistical noise even when it does not use all the available
data optimally and the column density distribution is adjusted using the
250\,$\mu$m observations only. The low noise can be understood as a result of
the strong intrinsic regularisation of the modelling procedure. In particular,
this precludes unphysical temperature variations (i.e., those greater than
allowed by the optical depths) and keeps the estimates reasonable even when the
signal goes to zero.

\section{Discussion} \label{sect:discussion}

We have examined different ways of estimating the column density
based on dust emission maps, especially using {\em Herschel}
data in the bands between 160\,$\mu$m and 500\,$\mu$m. The methods
A--C try to recover the column density at a resolution better than
the lowest resolution of the input maps. These aim at a resolution of
18$\arcsec$ (the resolution of the 250\,$\mu$m observations), a
factor of two better than the resolution of the 500\,$\mu$m data. The
radiative transfer models also were constructed in a way that results
in column density information at the same resolution.

In the tests with the extended emission (Sect.~\ref{sect:extended}),
Method B performed consistently better than Method A, the errors near
dense clumps being smaller by up to $\sim$50\% (see
Figs.\ref{fig:tff031_map} and \ref{fig:collins_map_BG}). The
difference remained clear even when estimates were compared at lower
resolution, 36.3$\arcsec$, corresponding to the resolution of the
500\,$\mu$m maps. Therefore, the difference is not limited to direct
effects of resolution. Only in the case of internal heating Method A
exhibited noticeably smaller bias. In those cases Method B was close
to the results that would be obtained if data at all wavelengths were
available at the same 18.3$\arcsec$ resolution. However, these
estimates are biased because of the line-of-sight temperature
variations that lead to overestimation of the dust temperature and
underestimation of the column density. By underestimating the
temperature variations that exist on small spatial scales, Method A
was actually closer to the true column density at the location of
internally heated clumps (see Figs.~\ref{fig:collins_map_SOU} and
\ref{fig:collins_profile}).

In Method B, there is some freedom to select the wavelengths that
are used to derive the estimates $N(500)$, $N(350)$, etc. For example,
if one assumes that SPIRE data give a more reliable picture of column
density on large scales (or that they are less biased by temperature
variations), one can base the $N(500)$ estimate on data between
250\,$\mu$m and 500\,$\mu$m only. The other terms could include also
the shorter wavelengths but, because in Method B these are high pass
filtered correction terms, they would not be sensitive to large scale
artefacts. For example, possible arteficial gradients or high pass
filtering of the PACS maps themselves would have only a limited impact
on the derived column density maps.

We also examined as Method C other linear combinations of the three
constituent terms of Method B, $N(500)$, $\left[ N(350) -
N(350\rightarrow 500) \right]$, and $\left[ N(250) - N(250\rightarrow
350) \right]$. By selecting optimal multipliers (Methods C and D), it
was possible to increase the correlation with the true column density
by a small but significant amount. The fact that those multipliers were
not very similar for the two examined cloud models suggests that this
may not be a viable method for general use. The best correspondence
with the true column density was obtained with multipliers that were of
different signs. This shows that, on large scales, the different biases
of the $N(250)$, $N(350)$, and $N(500)$ estimates have a significant effect
on the final errors.

The differences between the methods were accentuated in the optically
thick cores (Sect.~\ref{sect:cores}). For cores with $A_{V} \ga
50^{\rm m}$ or more, the central column densities can be underestimated
by several tens of percent. The strongest errors observed for both
Method A and B are a factor of three and the associated column density
peaks are hardly visible in the surface brightness maps, even at
250\,$\mu$m. The optimised linear combination of $N(250)$, $N(350)$, and
$N(500)$ improves the fit at the highest column densities, raising the
column density estimates by up to a factor of two. However, because the
errors behave in a very non-linear fashion (as a function of column
density), this increases the errors at the lower column densities. 

We also examined the possibility of making a non-linear combination of
the $N(250)$, $N(350)$, and $N(500)$ estimates. The bias depends on the
wavelengths used and we found, as expected, that $N(250) < N(350) <
N(500)$. The differences increase as a function of column density. As a
result, a non-linear combination of the estimates resulted in
significant improvement in the accuracy of the column density
predictions (see Fig.~\ref{fig:collins_NL}). It remains to be
established whether the parameters are be stable enough, so that the
method could be reliably applied to actual observations. The presence
of internal heating sources was already seen to eliminate much of the
systematic behaviour of $N(250)$, $N(350)$, and $N(500)$ relative to
each other.

\begin{figure}
\centering
\includegraphics[width=8.7cm]{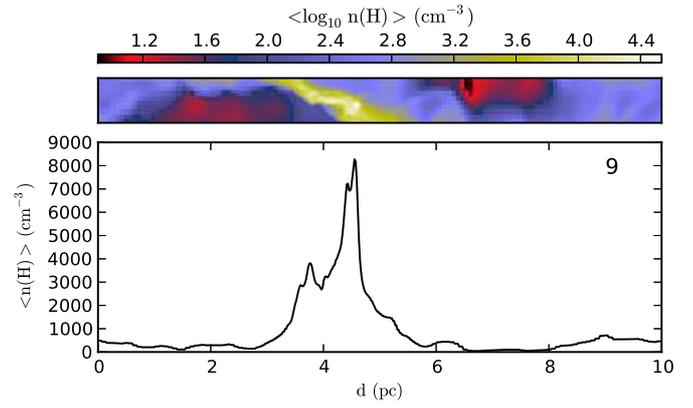}
\caption{
The line-of-sight structure of the field number 9. The data consist of the
density values within the modelled $5.3\arcmin \times 5.3\arcmin$ area, along
the full 10\,pc distance through the Cloud II cloud. The upper image shows that
density averaged over one direction perpendicular to the line-of-sight. The
lower plot shows the mean density as the function of line-of-sight distance.
The main structure is a filament with the long axis at $\sim$30
degree angle with respect to the line-of-sight.
}
\label{fig:n_los_39}
\end{figure}

\begin{figure}
\centering
\includegraphics[width=8.7cm]{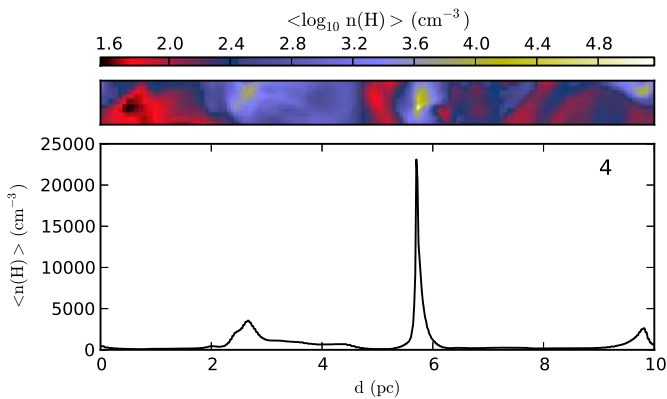}
\caption{
The line-of-sight structure of the field number 4 (see
Fig.~\ref{fig:n_los_39} for the details). The mass distribution is
dominated by a single, optically very thick core.
}
\label{fig:n_los_16}
\end{figure}

As the final method, we examined 3D radiative transfer modelling as a way to
estimate the column densities. In the case of Cloud II, this turned out to be
quite challenging because of the high optical depths. Together with the complex
density field this means that the radiation field illuminating the modelled
core could be strongly asymmetric. The lack of information about the
line-of-sight density structure is always a major problem and in this case, the
line-of-sight extent was more that ten times the perpendicular extent of the
modelled fields. Thus, also the radiation field could vary significantly along
this extent. The dense material was seen to be distributed over long distances.
This was in stark contrast with the assumed simple model where, for all lines
of sight, the density always peaked in the mid-plane. This maximises the
shadowing effect compared to the reality of isolated clumps
(Fig.~\ref{fig:n_los_16}) or oblique filaments (Fig.\ref{fig:n_los_39}).

In spite of these caveats, the modelling produced fair results. The model column
density was adjusted based on the 250\,$\mu$m observations and the attenuation of
the external field was adjusted according to the 160\,$\mu$m/500\,$\mu$m colour in
the central region. The accuracy of the results was typically slightly better than
for Method B. For the cores with a simple geometry (e.g., cores 1--3, see
Fig.~\ref{fig:9B}) the basic modelling produced very accurate density profiles,
while Method B underestimated the central column density by $\sim$10\%. For the
most opaque cores the modelling required the inclusion of an anisotropic
radiation field to avoid strong errors outside the central regions for which the
radiation field was tuned. Because the different structures along the line-of-sight
may be subjected to quite different radiation fields (e.g., of different intensity,
SED, and anisotropy), it may be difficult improve the results much further, at least
not without exhaustive examination of more complex models. The adjustment of the
width of the density distribution along the line-of-sight direction did not produce
very significant improvement. 

It is possible to construct models that (at least in the case of such
synthetic observations) reproduce all the observed surface brightness
maps to within the observational noise. However, in the case of Cloud
II this was already deemed too time consuming. The modelling
procedure used in this paper was very simple and, apart from the
column densities that were adjusted for each pixel separately, the
number of free parameters was small. As a result, the solution was
found in just some tens of iterations (run time of the order of one
hour per model). Even when the line-of-sight extent of the density
distribution was modified, all parameters could be updated relatively
independently using heuristic rules based on the observed and
modelled surface brightness maps. In more complex models (i.e., more
complex parameterisation of the cloud structure) the link between the
individual parameters and the surface brightness would be less
obvious and the solution would have to be obtained through general
optimisation. Depending on the number of parameters, this could be
orders of magnitude more time consuming. However, as long as the
models still exhibited systematic residuals (e.g.,
Fig.~\ref{fig:D16}), further improvements remain possible.

Cloud II is rather extreme in its opacity. The dense cores in nearby
molecular clouds would probably fall between Cloud II and the ellipsoidal cloud
of Sect.~\ref{sect:ellipsoid} in complexity (Fig.~\ref{fig:ellipsoid}).  For
the ellipsoidal cloud, if the external field was estimated correctly, the
modelling recovered the column density to within a couple of percent. If the
assumed ISRF was overestimated by a factor of two the errors remained below
$\sim$5\% and the signature of the wrong radiation field was visible in the
surface brightness maps. The same interpretation would be more difficult to
make in the case of real, irregularly shaped clouds. However, the results
suggest that for most of the clumps detected in nearby clouds one can, with
careful modelling, determine the column density profiles to an accuracy of a
few percent. Method B, possibly combined with a small bias correction, would
result in an almost similar accuracy and with considerably less effort. One
must also remember that we did not consider any of the uncertainties that are
related to dust properties and are likely to be the dominant errors in the
estimates of absolute column density.

\section{Conclusions} \label{sect:conclusions}

We have compared different, previously presented, methods to calculate
column density maps from dust emission, especially using the {\em
Herschel} wavelengths 160--500\,$\mu$m. Method A (Juvela et al.
(2012c) uses low resolution temperature estimates combined with higher
resolution intensity data. Method B \citep{Palmeirim2013} uses a
combination of column density estimates obtained using different
wavelength ranges. The methods try to recover the column density at a
resolution better than that of the lowest resolution input map. We
also test other modifications of the methods and compare these with
simple radiative transfer modelling that also is used to obtain the
column densities.

We have found that
\begin{itemize}

\item Both Method A and B give relatively reliable column density
estimates at the resolution of 250\,$\mu$m data while also making use
of the longer wavelengths.

\item By discarding temperature information on small scales, 
Method A shows greater errors for compact structures but is overall
less sensitive to noise.

\item When the examined clumps have internal heating sources,
Method B is consistent with results that would be obtained if high
resolution data were available at all wavelengths. However, these
underestimate the true column density and, because of favourable
cancellation of errors, Method A is sometimes closer to the true
column density.

\item Other linear combinations of the three terms of Method B can
increase the correlation by a small but significant amount. However,
this may not be a viable method for general use, as the multipliers
may depend on the cloud properties.

\item Radiative transfer modelling even with very simple
three-dimensional cloud models usually provides more accurate results.
However, the complexity of the models that are required for improved
results increases rapidly with the complexity and opacity of the
clouds.

\end{itemize}

\begin{acknowledgements}
The authors acknowledge the support of the Academy of Finland grant No. 250741.
TL acknowledges the support of the Academy of Finland grant No. 132291.
\end{acknowledgements}

\bibliography{biblio_v2.0}

\begin{thebibliography}{46}
\expandafter\ifx\csname natexlab\endcsname\relax\def\natexlab#1{#1}\fi

\bibitem[{{Andr{\'e}} {et~al.}(2010){Andr{\'e}}, {Men'shchikov}, {Bontemps},
  {K{\"o}nyves}, {Motte}, {Schneider}, {Didelon}, {Minier}, {Saraceno},
  {Ward-Thompson}, {di Francesco}, {White}, {Molinari}, {Testi}, {Abergel},
  {Griffin}, {Henning}, {Royer}, {Mer{\'{\i}}n}, {Vavrek}, {Attard},
  {Arzoumanian}, {Wilson}, {Ade}, {Aussel}, {Baluteau}, {Benedettini},
  {Bernard}, {Blommaert}, {Cambr{\'e}sy}, {Cox}, {di Giorgio}, {Hargrave},
  {Hennemann}, {Huang}, {Kirk}, {Krause}, {Launhardt}, {Leeks}, {Le Pennec},
  {Li}, {Martin}, {Maury}, {Olofsson}, {Omont}, {Peretto}, {Pezzuto}, {Prusti},
  {Roussel}, {Russeil}, {Sauvage}, {Sibthorpe}, {Sicilia-Aguilar}, {Spinoglio},
  {Waelkens}, {Woodcraft}, \& {Zavagno}}]{Andre2010}
{Andr{\'e}}, P., {Men'shchikov}, A., {Bontemps}, S., {et~al.} 2010, \aap, 518,
  L102

\bibitem[{{Andr\'e} {et~al.}(2000){Andr\'e}, {Ward-Thompson}, \&
  {Barsony}}]{Andre2000}
{Andr\'e}, P., {Ward-Thompson}, D., \& {Barsony}, M. 2000, Protostars and
  Planets IV, 59

\bibitem[{{Collins} {et~al.}(2011){Collins}, {Padoan}, {Norman}, \&
  {Xu}}]{Collins2011}
{Collins}, D.~C., {Padoan}, P., {Norman}, M.~L., \& {Xu}, H. 2011, \apj, 731,
  59

\bibitem[{{Collins} {et~al.}(2010){Collins}, {Xu}, {Norman}, {Li}, \&
  {Li}}]{Collins2010}
{Collins}, D.~C., {Xu}, H., {Norman}, M.~L., {Li}, H., \& {Li}, S. 2010, \apjs,
  186, 308

\bibitem[{{Compi{\`e}gne} {et~al.}(2011){Compi{\`e}gne}, {Verstraete}, {Jones},
  {Bernard}, {Boulanger}, {Flagey}, {Le Bourlot}, {Paradis}, \&
  {Ysard}}]{Compiegne2011}
{Compi{\`e}gne}, M., {Verstraete}, L., {Jones}, A., {et~al.} 2011, \aap, 525,
  A103

\bibitem[{{Draine}(2003)}]{Draine2003}
{Draine}, B.~T. 2003, \apj, 598, 1017

\bibitem[{{Evans} {et~al.}(2001){Evans}, {Rawlings}, {Shirley}, \&
  {Mundy}}]{Evans2001}
{Evans}, II, N.~J., {Rawlings}, J.~M.~C., {Shirley}, Y.~L., \& {Mundy}, L.~G.
  2001, \apj, 557, 193

\bibitem[{{Foster} \& {Goodman}(2006)}]{Foster2006}
{Foster}, J.~B. \& {Goodman}, A.~A. 2006, \apjl, 636, L105

\bibitem[{{Goodman} {et~al.}(2009){Goodman}, {Pineda}, \&
  {Schnee}}]{Goodman2009}
{Goodman}, A.~A., {Pineda}, J.~E., \& {Schnee}, S.~L. 2009, \apj, 692, 91

\bibitem[{{Griffin} {et~al.}(2010){Griffin}, {Abergel}, {Abreu}, {Ade},
  {Andr{\'e}}, {Augueres}, {Babbedge}, {Bae}, {Baillie}, {Baluteau}, {Barlow},
  {Bendo}, {Benielli}, {Bock}, {Bonhomme}, {Brisbin}, {Brockley-Blatt},
  {Caldwell}, {Cara}, {Castro-Rodriguez}, {Cerulli}, {Chanial}, {Chen},
  {Clark}, {Clements}, {Clerc}, {Coker}, {Communal}, {Conversi}, {Cox},
  {Crumb}, {Cunningham}, {Daly}, {Davis}, {de Antoni}, {Delderfield}, {Devin},
  {di Giorgio}, {Didschuns}, {Dohlen}, {Donati}, {Dowell}, {Dowell}, {Duband},
  {Dumaye}, {Emery}, {Ferlet}, {Ferrand}, {Fontignie}, {Fox}, {Franceschini},
  {Frerking}, {Fulton}, {Garcia}, {Gastaud}, {Gear}, {Glenn}, {Goizel},
  {Griffin}, {Grundy}, {Guest}, {Guillemet}, {Hargrave}, {Harwit}, {Hastings},
  {Hatziminaoglou}, {Herman}, {Hinde}, {Hristov}, {Huang}, {Imhof}, {Isaak},
  {Israelsson}, {Ivison}, {Jennings}, {Kiernan}, {King}, {Lange}, {Latter},
  {Laurent}, {Laurent}, {Leeks}, {Lellouch}, {Levenson}, {Li}, {Li},
  {Lilienthal}, {Lim}, {Liu}, {Lu}, {Madden}, {Mainetti}, {Marliani}, {McKay},
  {Mercier}, {Molinari}, {Morris}, {Moseley}, {Mulder}, {Mur}, {Naylor},
  {Nguyen}, {O'Halloran}, {Oliver}, {Olofsson}, {Olofsson}, {Orfei}, {Page},
  {Pain}, {Panuzzo}, {Papageorgiou}, {Parks}, {Parr-Burman}, {Pearce},
  {Pearson}, {P{\'e}rez-Fournon}, {Pinsard}, {Pisano}, {Podosek}, {Pohlen},
  {Polehampton}, {Pouliquen}, {Rigopoulou}, {Rizzo}, {Roseboom}, {Roussel},
  {Rowan-Robinson}, {Rownd}, {Saraceno}, {Sauvage}, {Savage}, {Savini},
  {Sawyer}, {Scharmberg}, {Schmitt}, {Schneider}, {Schulz}, {Schwartz},
  {Shafer}, {Shupe}, {Sibthorpe}, {Sidher}, {Smith}, {Smith}, {Smith},
  {Spencer}, {Stobie}, {Sudiwala}, {Sukhatme}, {Surace}, {Stevens}, {Swinyard},
  {Trichas}, {Tourette}, {Triou}, {Tseng}, {Tucker}, {Turner}, {Vaccari},
  {Valtchanov}, {Vigroux}, {Virique}, {Voellmer}, {Walker}, {Ward}, {Waskett},
  {Weilert}, {Wesson}, {White}, {Whitehouse}, {Wilson}, {Winter}, {Woodcraft},
  {Wright}, {Xu}, {Zavagno}, {Zemcov}, {Zhang}, \& {Zonca}}]{Griffin2010}
{Griffin}, M.~J., {Abergel}, A., {Abreu}, A., {et~al.} 2010, \aap, 518, L3

\bibitem[{{Juvela}(2005)}]{Juvela2005}
{Juvela}, M. 2005, \aap, 440, 531

\bibitem[{{Juvela} {et~al.}(2012{\natexlab{a}}){Juvela}, {Malinen}, \&
  {Lunttila}}]{Juvela2012_mhdfil}
{Juvela}, M., {Malinen}, J., \& {Lunttila}, T. 2012{\natexlab{a}}, \aap, 544,
  A141

\bibitem[{{Juvela} \& {Padoan}(2003)}]{Juvela2003}
{Juvela}, M. \& {Padoan}, P. 2003, \aap, 397, 201

\bibitem[{{Juvela} {et~al.}(2008){Juvela}, {Pelkonen}, {Padoan}, \&
  {Mattila}}]{Juvela2008}
{Juvela}, M., {Pelkonen}, V.-M., {Padoan}, P., \& {Mattila}, K. 2008, \aap,
  480, 445

\bibitem[{{Juvela} {et~al.}(2012{\natexlab{b}}){Juvela}, {Pelkonen}, {White},
  {K{\"o}nyves}, {Kirk}, \& {Andr{\'e}}}]{Juvela2012_CrA_III}
{Juvela}, M., {Pelkonen}, V.-M., {White}, G.~J., {et~al.} 2012{\natexlab{b}},
  \aap, 544, A14

\bibitem[{{Juvela} {et~al.}(2012{\natexlab{c}}){Juvela}, {Ristorcelli},
  {Pagani}, {Doi}, {Pelkonen}, {Marshall}, {Bernard}, {Falgarone}, {Malinen},
  {Marton}, {McGehee}, {Montier}, {Motte}, {Paladini}, {T{\'o}th}, {Ysard},
  {Zahorecz}, \& {Zavagno}}]{Juvela2012_GCC_III}
{Juvela}, M., {Ristorcelli}, I., {Pagani}, L., {et~al.} 2012{\natexlab{c}},
  \aap, 541, A12

\bibitem[{{Juvela} \& {Ysard}(2012{\natexlab{a}})}]{Juvela2012_chi2}
{Juvela}, M. \& {Ysard}, N. 2012{\natexlab{a}}, \aap, 541, A33

\bibitem[{{Juvela} \& {Ysard}(2012{\natexlab{b}})}]{Juvela2012_TB}
{Juvela}, M. \& {Ysard}, N. 2012{\natexlab{b}}, \aap, 539, A71

\bibitem[{{Lehtinen} \& {Mattila}(1996)}]{Lehtinen1996}
{Lehtinen}, K. \& {Mattila}, K. 1996, \aap, 309, 570

\bibitem[{{Lombardi} {et~al.}(2006){Lombardi}, {Alves}, \&
  {Lada}}]{Lombardi2006}
{Lombardi}, M., {Alves}, J., \& {Lada}, C.~J. 2006, \aap, 454, 781

\bibitem[{{Lunttila} \& {Juvela}(2012)}]{Lunttila2012}
{Lunttila}, T. \& {Juvela}, M. 2012, \aap, 544, A52

\bibitem[{{Malinen} {et~al.}(2011){Malinen}, {Juvela}, {Collins}, {Lunttila},
  \& {Padoan}}]{Malinen2011}
{Malinen}, J., {Juvela}, M., {Collins}, D.~C., {Lunttila}, T., \& {Padoan}, P.
  2011, \aap, 530, A101+

\bibitem[{{Malinen} {et~al.}(2013){Malinen}, {Juvela}, M., \&
  {Pelkonen}}]{Malinen2013}
{Malinen}, J., {Juvela}, M., M., R., \& {Pelkonen}, V.-M. 2013, \aap, in
  preparation

\bibitem[{{Malinen} {et~al.}(2012){Malinen}, {Juvela}, {Rawlings},
  {Ward-Thompson}, {Palmeirim}, \& {Andr{\'e}}}]{Malinen2012}
{Malinen}, J., {Juvela}, M., {Rawlings}, M.~G., {et~al.} 2012, \aap, 544, A50

\bibitem[{{Mathis} {et~al.}(1983){Mathis}, {Mezger}, \& {Panagia}}]{Mathis1983}
{Mathis}, J.~S., {Mezger}, P.~G., \& {Panagia}, N. 1983, \aap, 128, 212

\bibitem[{{Meny} {et~al.}(2007){Meny}, {Gromov}, {Boudet}, {Bernard},
  {Paradis}, \& {Nayral}}]{Meny2007}
{Meny}, C., {Gromov}, V., {Boudet}, N., {et~al.} 2007, \aap, 468, 171

\bibitem[{{Molinari} {et~al.}(2010){Molinari}, {Swinyard}, {Bally}, {Barlow},
  {Bernard}, {Martin}, {Moore}, {Noriega-Crespo}, {Plume}, {Testi}, {Zavagno},
  {Abergel}, {Ali}, {Anderson}, {Andr{\'e}}, {Baluteau}, {Battersby},
  {Beltr{\'a}n}, {Benedettini}, {Billot}, {Blommaert}, {Bontemps}, {Boulanger},
  {Brand}, {Brunt}, {Burton}, {Calzoletti}, {Carey}, {Caselli}, {Cesaroni},
  {Cernicharo}, {Chakrabarti}, {Chrysostomou}, {Cohen}, {Compiegne}, {de
  Bernardis}, {de Gasperis}, {di Giorgio}, {Elia}, {Faustini}, {Flagey},
  {Fukui}, {Fuller}, {Ganga}, {Garcia-Lario}, {Glenn}, {Goldsmith}, {Griffin},
  {Hoare}, {Huang}, {Ikhenaode}, {Joblin}, {Joncas}, {Juvela}, {Kirk},
  {Lagache}, {Li}, {Lim}, {Lord}, {Marengo}, {Marshall}, {Masi}, {Massi},
  {Matsuura}, {Minier}, {Miville-Desch{\^e}nes}, {Montier}, {Morgan}, {Motte},
  {Mottram}, {M{\"u}ller}, {Natoli}, {Neves}, {Olmi}, {Paladini}, {Paradis},
  {Parsons}, {Peretto}, {Pestalozzi}, {Pezzuto}, {Piacentini}, {Piazzo},
  {Polychroni}, {Pomar{\`e}s}, {Popescu}, {Reach}, {Ristorcelli}, {Robitaille},
  {Robitaille}, {Rod{\'o}n}, {Roy}, {Royer}, {Russeil}, {Saraceno}, {Sauvage},
  {Schilke}, {Schisano}, {Schneider}, {Schuller}, {Schulz}, {Sibthorpe},
  {Smith}, {Smith}, {Spinoglio}, {Stamatellos}, {Strafella}, {Stringfellow},
  {Sturm}, {Taylor}, {Thompson}, {Traficante}, {Tuffs}, {Umana}, {Valenziano},
  {Vavrek}, {Veneziani}, {Viti}, {Waelkens}, {Ward-Thompson}, {White},
  {Wilcock}, {Wyrowski}, {Yorke}, \& {Zhang}}]{Molinari2010}
{Molinari}, S., {Swinyard}, B., {Bally}, J., {et~al.} 2010, \aap, 518, L100

\bibitem[{{Motte} {et~al.}(1998){Motte}, {Andre}, \& {Neri}}]{Motte1998}
{Motte}, F., {Andre}, P., \& {Neri}, R. 1998, A\&A, 336, 150

\bibitem[{{Nakajima} {et~al.}(2008){Nakajima}, {Kandori}, {Tamura}, {Nagata},
  {Sato}, \& {Sugitani}}]{Nakajima2008}
{Nakajima}, Y., {Kandori}, R., {Tamura}, M., {et~al.} 2008, \pasj, 60, 731

\bibitem[{{Nakajima} {et~al.}(2003){Nakajima}, {Nagata}, {Sato}, {Nagayama},
  {Nagashima}, {Kato}, {Kurita}, {Kawai}, {Tamura}, {Nakaya}, \&
  {Sugitani}}]{Nakajima2003}
{Nakajima}, Y., {Nagata}, T., {Sato}, S., {et~al.} 2003, \aj, 125, 1407

\bibitem[{{Nielbock} {et~al.}(2012){Nielbock}, {Launhardt}, {Steinacker},
  {Stutz}, {Balog}, {Beuther}, {Bouwman}, {Henning}, {Hily-Blant},
  {Kainulainen}, {Krause}, {Linz}, {Lippok}, {Ragan}, {Risacher}, \&
  {Schmiedeke}}]{Nielbock2012}
{Nielbock}, M., {Launhardt}, R., {Steinacker}, J., {et~al.} 2012, \aap, 547,
  A11

\bibitem[{{Ossenkopf} \& {Henning}(1994)}]{Ossenkopf1994}
{Ossenkopf}, V. \& {Henning}, T. 1994, A\&A, 291, 943

\bibitem[{{Padoan} \& {Nordlund}(2011)}]{PadoanNordlund2011}
{Padoan}, P. \& {Nordlund}, {\AA}. 2011, \apj, 730, 40

\bibitem[{{Pagani} {et~al.}(2010){Pagani}, {Steinacker}, {Bacmann}, {Stutz}, \&
  {Henning}}]{Pagani2010}
{Pagani}, L., {Steinacker}, J., {Bacmann}, A., {Stutz}, A., \& {Henning}, T.
  2010, Science, 329, 1622

\bibitem[{{Palmeirim} {et~al.}(2013){Palmeirim}, {Andr{\'e}}, {Kirk},
  {Ward-Thompson}, {Arzoumanian}, {K{\"o}nyves}, {Didelon}, {Schneider},
  {Benedettini}, {Bontemps}, {Di Francesco}, {Elia}, {Griffin}, {Hennemann},
  {Hill}, {Martin}, {Men'shchikov}, {Molinari}, {Motte}, {Nguyen Luong},
  {Nutter}, {Peretto}, {Pezzuto}, {Roy}, {Rygl}, {Spinoglio}, \&
  {White}}]{Palmeirim2013}
{Palmeirim}, P., {Andr{\'e}}, P., {Kirk}, J., {et~al.} 2013, \aap, 550, A38

\bibitem[{{Pilbratt} {et~al.}(2010){Pilbratt}, {Riedinger}, {Passvogel},
  {Crone}, {Doyle}, {Gageur}, {Heras}, {Jewell}, {Metcalfe}, {Ott}, \&
  {Schmidt}}]{Pilbratt2010}
{Pilbratt}, G.~L., {Riedinger}, J.~R., {Passvogel}, T., {et~al.} 2010, \aap,
  518, L1

\bibitem[{{Poglitsch} {et~al.}(2010){Poglitsch}, {Waelkens}, {Geis},
  {Feuchtgruber}, {Vandenbussche}, {Rodriguez}, {Krause}, {Renotte}, {van
  Hoof}, {Saraceno}, {Cepa}, {Kerschbaum}, {Agn{\`e}se}, {Ali}, {Altieri},
  {Andreani}, {Augueres}, {Balog}, {Barl}, {Bauer}, {Belbachir}, {Benedettini},
  {Billot}, {Boulade}, {Bischof}, {Blommaert}, {Callut}, {Cara}, {Cerulli},
  {Cesarsky}, {Contursi}, {Creten}, {De Meester}, {Doublier}, {Doumayrou},
  {Duband}, {Exter}, {Genzel}, {Gillis}, {Gr{\"o}zinger}, {Henning},
  {Herreros}, {Huygen}, {Inguscio}, {Jakob}, {Jamar}, {Jean}, {de Jong},
  {Katterloher}, {Kiss}, {Klaas}, {Lemke}, {Lutz}, {Madden}, {Marquet},
  {Martignac}, {Mazy}, {Merken}, {Montfort}, {Morbidelli}, {M{\"u}ller},
  {Nielbock}, {Okumura}, {Orfei}, {Ottensamer}, {Pezzuto}, {Popesso},
  {Putzeys}, {Regibo}, {Reveret}, {Royer}, {Sauvage}, {Schreiber}, {Stegmaier},
  {Schmitt}, {Schubert}, {Sturm}, {Thiel}, {Tofani}, {Vavrek}, {Wetzstein},
  {Wieprecht}, \& {Wiezorrek}}]{Poglitsch2010}
{Poglitsch}, A., {Waelkens}, C., {Geis}, N., {et~al.} 2010, \aap, 518, L2

\bibitem[{{Ridderstad} \& {Juvela}(2010)}]{Ridderstad2010}
{Ridderstad}, M. \& {Juvela}, M. 2010, \aap, 520, A18+

\bibitem[{{Schneider} {et~al.}(2011){Schneider}, {Bontemps}, {Simon},
  {Ossenkopf}, {Federrath}, {Klessen}, {Motte}, {Andr{\'e}}, {Stutzki}, \&
  {Brunt}}]{Schneider2011}
{Schneider}, N., {Bontemps}, S., {Simon}, R., {et~al.} 2011, \aap, 529, A1+

\bibitem[{{Shetty} {et~al.}(2009{\natexlab{a}}){Shetty}, {Kauffmann}, {Schnee},
  \& {Goodman}}]{Shetty2009b}
{Shetty}, R., {Kauffmann}, J., {Schnee}, S., \& {Goodman}, A.~A.
  2009{\natexlab{a}}, \apj, 696, 676

\bibitem[{{Shetty} {et~al.}(2009{\natexlab{b}}){Shetty}, {Kauffmann}, {Schnee},
  {Goodman}, \& {Ercolano}}]{Shetty2009a}
{Shetty}, R., {Kauffmann}, J., {Schnee}, S., {Goodman}, A.~A., \& {Ercolano},
  B. 2009{\natexlab{b}}, \apj, 696, 2234

\bibitem[{{Stamatellos} \& {Whitworth}(2003)}]{StamatellosWhitworth2003}
{Stamatellos}, D. \& {Whitworth}, A.~P. 2003, \aap, 407, 941

\bibitem[{{Steinacker} {et~al.}(2010){Steinacker}, {Pagani}, {Bacmann}, \&
  {Guieu}}]{Steinacker2010}
{Steinacker}, J., {Pagani}, L., {Bacmann}, A., \& {Guieu}, S. 2010, \aap, 511,
  A9+

\bibitem[{{Stepnik} {et~al.}(2003){Stepnik}, {Abergel}, {Bernard}, {Boulanger},
  {Cambr{\'e}sy}, {Giard}, {Jones}, {Lagache}, {Lamarre}, {Meny}, {Pajot}, {Le
  Peintre}, {Ristorcelli}, {Serra}, \& {Torre}}]{Stepnik2003}
{Stepnik}, B., {Abergel}, A., {Bernard}, J., {et~al.} 2003, \aap, 398, 551

\bibitem[{{Wilcock} {et~al.}(2012){Wilcock}, {Ward-Thompson}, {Kirk},
  {Stamatellos}, {Whitworth}, {Battersby}, {Elia}, {Fuller}, {DiGiorgio},
  {Griffin}, {Molinari}, {Martin}, {Mottram}, {Peretto}, {Pestalozzi},
  {Schisano}, {Smith}, \& {Thompson}}]{Wilcock2012}
{Wilcock}, L.~A., {Ward-Thompson}, D., {Kirk}, J.~M., {et~al.} 2012, \mnras,
  424, 716

\bibitem[{{Ysard} {et~al.}(2012){Ysard}, {Juvela}, {Demyk}, {Guillet},
  {Abergel}, {Bernard}, {Malinen}, {M{\'e}ny}, {Montier}, {Paradis},
  {Ristorcelli}, \& {Verstraete}}]{YsardJuvela2012}
{Ysard}, N., {Juvela}, M., {Demyk}, K., {et~al.} 2012, \aap, 542, A21

\end{thebibliography}

\end{document}